\documentclass[12pt]{article}
\usepackage{psfrag}
\usepackage{amsmath}
\usepackage{amsthm}
\usepackage{amssymb}
\usepackage{epsfig}
\usepackage{euscript}
\usepackage{array}
\usepackage{bbold}
\usepackage{cancel}
\usepackage{mathtools}
\usepackage{empheq}
\usepackage{graphicx}
\usepackage{subfigure}
\usepackage{upgreek}
\usepackage{epstopdf}
\usepackage{booktabs}
\usepackage{mathrsfs}
\usepackage{amsbsy}

\usepackage{hyperref}
\usepackage{verbatim}

\theoremstyle{definition}

\theoremstyle{remark}

\newcounter{multieqs}

\newcommand{\be}{\begin{equation}}
\newcommand{\ee}{\end{equation}}
\newcommand{\eq}[1]{(\ref{#1})}
\newcommand{\bit}{\begin{itemize}}  \newcommand{\eit}{\end{itemize}}
\newcommand{\ben}{\begin{enumerate}}  \newcommand{\een}{\end{enumerate}}

\newcommand{\ket}[1]{|#1 \rangle}

\newcommand{\bm}[1]{\mbox{\boldmath $#1$}}
\newcommand{\rf}[1]{(\ref{#1})}

\def\bd{\begin{document}}
\def\ed{\end{document}}
 \def\bea{\begin{eqnarray}}
 \def\eea{\end{eqnarray}}
\let\bm=\bibitem

\def\la{\langle}
\def\ra{\rangle}

\def\npb#1#2#3{Nucl. Phys. {\bf{B#1}} #3 (#2)}
\def\plb#1#2#3{Phys. Lett. {\bf{#1B}} #3 (#2)}
\def\prl#1#2#3{Phys. Rev. Lett. {\bf{#1}} #3 (#2)}
\def\prd#1#2#3{Phys. Rev. {D bf{#1}} #3 (#2)}
\def\cmp#1#2#3{Comm. Math. Phys. {\bf{#1}} #3 (#2)}
\def\cqg#1#2#3{Class. Quantum Grav. {\bf{#1}} #3 (#2)}
\def\nppsa#1#2#3{Nucl. Phys. B (Proc. Suppl.) {\bf{#1A}}#3 (#2)}
\def\ap#1#2#3{Ann. of Phys. {\bf{#1}} #3 (#2)}
\def\ijmp#1#2#3{Int. J. Mod. Phys. {\bf{A#1}} #3 (#2)}
\def\rmp#1#2#3{Rev. Mod. Phys. {\bf{#1}} #3 (#2)}
\def\mpla#1#2#3{Mod. Phys. Lett. {\bf A#1} #3 (#2)}
\def\jhep#1#2#3{J. High Energy Phys. {\bf #1} #3 (#2)}
\def\atmp#1#2#3{Adv. Theor. Math. Phys. {\bf #1} #3 (#2)}

\def\N{{\cal N}}
\def\sst{\scriptscriptstyle}
\def\thetabar{\bar\theta}
\def\Tr{{\rm Tr}}
\def\one{\mbox{1 \kern-.59em {\rm l}}}

\def\a{\alpha}      \def\da{{\dot\alpha}}  \def\dA{{\dot A}}
\def\b{\beta}       \def\db{{\dot\beta}}
\def\g{\gamma}  \def\G{\Gamma}  \def\dc{{\dot\gamma}}
\def\d{\delta}  \def\D{\Delta}  \def\ddt{\dot\delta}
\def\e{\epsilon}
\def\ve{\varepsilon}
\def\uve{\upvarepsilon}
\def\f{\phi}    \def\F{\Phi}    \def\vvf{\f}
\def\vphi{\varphi}
\def\h{\eta}
\def\k{\kappa}
\def\l{\lambda} \def\L{\Lambda}
\def\m{\mu} \def\n{\nu}
\def\o{\omega}
\def\p{\pi} \def\P{\Pi}
\def\r{\rho}
\def\s{\sigma}  \def\S{\Sigma}
\def\t{\tau}
\def\th{\theta} \def\Th{\Theta} \def\vth{\vartheta}
\def\X{\Xeta}
\def\z{\zeta}

\def\na{\nabla}

\def\cA{{\mathscr A}} \def\cB{{\cal B}} \def\cC{{\cal C}}
\def\cD{{\cal D}} \def\cE{{\cal E}} \def\cF{{\cal F}}
\def\cG{{\cal G}} \def\cH{{\cal H}} \def\cI{{\cal I}}
\def\cJ{{\mathscr J}} \def\cK{{\cal K}} \def\cL{{\cal L}}
\def\cM{{\cal M}} \def\cN{{\cal N}} \def\cO{{\cal O}}
\def\cP{{\cal P}} \def\cQ{{\cal Q}} \def\cR{{\cal R}}
\def\cS{{\cal S}} \def\cT{{\cal T}} \def\cU{{\cal U}}
\def\cV{{\cal V}} \def\cW{{\cal W}} \def\cX{{\cal X}}
\def\cY{{\cal Y}} \def\cZ{{\cal Z}}

\def\ua{\underline{\alpha}}
\def\uc{\underline{\phantom{\alpha}}\!\!\!\gamma}
\def\um{\underline{\mu}}
\def\ud{\underline\delta}
\def\ue{\underline\epsilon}
\def\una{\underline a}\def\unA{\underline A}
\def\unb{\underline b}\def\unB{\underline B}
\def\unc{\underline c}\def\unC{\underline C}
\def\und{\underline d}\def\unD{\underline D}
\def\une{\underline e}\def\unE{\underline E}
\def\unf{\underline{\phantom{e}}\!\!\!\! f}\def\unF{\underline F}
\def\unm{\underline m}\def\unM{{\underline M}}
\def\unn{\underline n}\def\unN{{\underline N}}
\def\unp{\underline{\phantom{a}}\!\!\! p}\def\unP{\underline P}
\def\unq{\underline{\phantom{a}}\!\!\! q}
\def\unQ{\underline{\phantom{A}}\!\!\!\! Q}
\def\unH{\underline{H}}

\def\As {{A \hspace{-6.4pt} \slash}\;}
\def\bs {{b \hspace{-6.4pt} \slash}\;}
\def\Ds {{D \hspace{-6.4pt} \slash}\;}
\def\Gts {{\Gt \hspace{-6.4pt} \slash}\;}
\def\ds {{\del \hspace{-6.4pt} \slash}\;}
\def\ss {{\s \hspace{-6.4pt} \slash}\;}
\def\ks {{ k \hspace{-6.4pt} \slash}\;}
\def\ps {{p \hspace{-6.4pt} \slash}\;}
\def\xs {{x \hspace{-6.4pt} \slash}\;}
\def\pas {{{p_1} \hspace{-6.4pt} \slash}\;}
\def\pbs {{{p_2} \hspace{-6.4pt} \slash}\;}
\def\cFs {{{\cal F} \hspace{-6.4pt} \slash}\;}
\def\Dss {{D \hspace{-7.5pt} \slash}\;}
\def\dss {{\del \hspace{-7.0pt} \slash}\;}

\def\Ah{{\hat{A}}}
\def\Ch{{\hat{C}}}
\def\Dh{{\hat{D}}}
\def\Gh{{\hat{G}}}
\def\Fh{{\hat{F}}}
\def\Ih{{\hat{I}}}
\def\Jh{{\hat{J}}}
\def\Kh{{\hat{K}}}
\def\Lh{{\hat{L}}}
\def\Ph{{\hat{P}}}
\def\Rh{{\hat{R}}}
\def\Vh{{\hat{V}}}
\def\Xh{{\hat{X}}}

\def\ah{{\hat{\a}}}
\def\bh{{\hat{\b}}}
\def\gh{{\hat{\g}}}
\def\dh{{\hat{\d}}}
\def\rh{{\hat{\r}}}
\def\hh{\hat{h}}
\def\uh{\hat{u}}
\def\xh{\hat{x}}
\def\yh{\hat{y}}
\def\ph{\hat{p}}
\def\xih{\hat{\xi}}
\def\chih{\hat{\chi}}
\def\Psih{\hat{\Psi}}
\def\phih{\hat{\phi}}

\def\psit{\tilde{\psi}}
\def\Psit{\tilde{\Psi}}
\def\Psibt{\tilde{\bar{Psi}}}

\def\st{\tilde{\sigma}}

\def\delt{\tilde{\delta}}
\def\Phit{\tilde{\Phi}}
\def\Phitb{\overline{\tilde{Phi}}}
\def\tht{\tilde{\th}}
\def\lt{\tilde{\l}}
\def\chit{\tilde{\chi}}
\def\phit{\tilde{\phi}}

\def\At{\tilde{A}}
\def\Bt{\tilde{B}}
\def\Ct{\tilde{C}}
\def\Dt{\tilde{D}}
\def\Et{\tilde{E}}
\def\Ft{\tilde{F}}
\def\Gt{\tilde{G}}
\def\Ht{\tilde{H}}
\def\It{\tilde{I}}
\def\Jt{\tilde{J}}
\def\Qt{\tilde{Q}}
\def\Rt{\tilde{R}}
\def\Mt{\tilde{M }}
\def\Nt{\tilde{N}}
\def\St{\tilde{S}}
\def\Vt{\tilde{V}}
\def\Xt{\tilde{X}}
\def\at{\tilde{a}}
\def\ct{\tilde{c}}
\def\dt{\tilde{d}}
\def\htt{\tilde{h}}
\def\ft{\tilde{f}}
\def\gt{\tilde{g}}
\def\pt{\tilde{p}}
\def\qt{\tilde{q}}
\def\vt{\tilde{v}}
\def\nt{\tilde{n}}
\def\ut{\tilde{u}}
\def\wt{\tilde{w}}
\def\zt{\tilde{z}}
\def\xt{\tilde{x}}
\def\yt{\tilde{y}}
\def\Psit{\tilde{\Psi}}
\def\vphit{\tilde{\varphi}}
\def\tD{\tilde{\D}}

\def\eb{\bar{\epsilon}}
\def\delb{\bar{\partial}}
\def\thb{\bar{\theta}}
\def\mub{\bar{\mu}}
\def\lamb{\bar{\l}}
\def\psib{\bar{\psi}}
\def\sb{\bar{\sigma}}
\def\xib{\bar{\xi}}
\def\chib{\bar{\chi}}

\def\Psib{\bar{\Psi}}
\def\Phib{\bar{\Phi}}
\def\Lamb{\bar{\Lambda}}
\def\Sb{{\overline \Sigma}}
\def\cb{\bar{c}}
\def\hb{\bar{h}}
\def\qb{\bar{q}}
\def\wb{\bar{w}}
\def\ub{\bar{u}}
\def\zb{{\bar{z}}}
\def\Hb{\bar{H}}
\def\Qb{{\bar Q}}
\def\Omegab{\overline{\Omega}}
\def\ob{\overline{\omega}}

\def\Ab{{\overline A}} \def\Bb{{\overline B}} \def\Cb{{\overline C}}
\def\Db{{\overline D}} \def\Eb{{\overline E}} \def\Fb{{\overline F}}
\def\Gb{{\overline G}}
\def\Ib{{\overline I}}
\def\Jb{{\overline J}} \def\Kb{{\overline K}} \def\Lb{{\overline L}}
\def\Mb{{\overline M}} \def\Nb{{\overline N}} \def\Ob{{\overline O}}
\def\Pb{{\overline P}}  \def\Rb{{\overline R}}
 \def\Tb{{\overline T}} \def\Ub{{\overline U}}
\def\Vb{{\overline V}} \def\Wb{{\overline W}} \def\Xb{{\overline X}}
\def\Yb{{\overline Y}} \def\Zb{{\overline Z}}

\def\fb{{\overline f}}
\def\gb{{\overline g}}
\def\mb{{\overline m}}
\def\lb{{\overline l}}
\def\yb{{\overline y}}

\def\ba{{\bf a}}
\def\bk{{\bf k}}
\def\bl{{\bf l}}
\def\bp{{\bf p}}
\def\bq{{\bf q}}
\def\br{{\bf r}}
\def\bt{{\bf t}}
\def\bu{{\bf u}}
\def\bv{{\bf v}}
\def\bx{{\bf x}}
\def\by{{\bf y}}
\def\bA{{\bf A}}
\def\bB{{\bf B}}
\def\bR{{\bf R}}
\def\bV{{\bf V}}

\def\bz{{\boldsymbol{\zeta}}}

\def\bone{{\bf 1}}

\def\va{{\vec a}}
\def\vk{{\vec k}}
\def\vp{{\vec p}}
\def\vq{{\vec q}}
\def\vx{{\vec x}}
\def\vy{{\vec y}}
\def\vu{{\vec u}}
\def\vv{{\vec v}}
\def \vH{{\vec H}}
\def \vg{{\vec g}}

\def\vs{{\vec \sigma}}
\def\vtau{{\vec \tau}}

\def\frA{\mathfrak{A}}
\def\frB{\mathfrak{B}}
\def\frC{\mathfrak{C}}
\def\frD{\mathfrak{D}}
\def\frE{\mathfrak{E}}
\def\frF{\mathfrak{F}}
\def\frG{\mathfrak{G}}
\def\frH{\mathfrak{H}}
\def\frM{\mathfrak{M}}
\def\frN{\mathfrak{N}}
\def\frR{\mathfrak{R}}
\def\frW{\mathfrak{W}}

\def\fra{\mathfrak{a}}
\def\frb{\mathfrak{b}}
\def\frf{\mathfrak{f}}
\def\frg{\mathfrak{g}}
\def\frh{\mathfrak{h}}
\def\frl{\mathfrak{l}}
\def\frs{\mathfrak{s}}
\def\fri{\mathfrak{i}}
\def\frj{\mathfrak{j}}

\def\ma{\mathfrak{a}}
\def\mg{\mathfrak{g}}
\def\mh{\mathfrak{h}}
\def\mR{\mathfrak{R}}
\def\mN{\mathfrak{N}}

\newcommand{\nn}{{\nonumber}}

\def\d{\delta}\def\D{\Delta}\def\ddt{\dot\delta}

\def\pa{\partial} \def\del{\partial}
\def\xx{\times}
\def\uno{\mbox{1 \kern-.59em {\rm l}}}

\def\trp{^{\top}}
\def\inv{^{-1}}
\def\dag{\dagger}
\def\pr{^{\prime}}

\def\rar{\rightarrow}
\def\lar{\leftarrow}
\def\lrar{\leftrightarrow}

\newcommand{\0}{\,\!}      
\def\one{1\!\!1\,\,}
\def\im{\imath}
\def\jm{\jmath}

\newcommand{\tr}{\mbox{tr}}
\newcommand{\slsh}[1]{/ \!\!\!\! #1}

\def\vac{|0\rangle}
\def\lvac{\langle 0|}

\def\hlf{\frac{1}{2}}
\def\ove#1{\frac{1}{#1}}
\newcommand{\hot}[1]{\frac{#1}{2}}

\def\Box{\square}
\def\CC {\mathbb{C}}
\def\FF {\mathbb{F}}
\def\RR{\mathbb{R}}
\def\NN{\mathbb{N}}
\def\ZZ{\mathbb{Z}}
\def\bb#1{{\bf #1}}
\def\bcomment#1{}
\def\bfhat#1{{\bf \hat{#1}}}
\def\VEV#1{\left\langle #1\right\rangle}

\newcommand{\ex}[1]{{\rm e}^{#1}} \def\ii{{\rm i}}

\newcommand{\lrbrk}[1]{\left(#1\right)}
\newcommand{\lrsbrk}[1]{\left[#1\right]}
\newcommand{\sfrac}[2]{{\textstyle\frac{#1}{#2}}}

\def\stw{{\sqrt{2}}}

\def\rf {{\rm f}}
\def\ri {{\rm i}}
\def\rj {{\rm j}}
\def\rn {{\rm n}}
\def\rk {{\rm k}}
\def\rl {{\rm l}}
\def\rr {{\rm r}}
\def\rs {{\scriptscriptstyle \rm S}}
\def\rt {{\scriptscriptstyle \rm T}}

\def\rQ {{\scriptscriptstyle \rm \cQ}}
\def\rR {{\scriptscriptstyle \rm \cR}}

\def\cQb{{\cal \Qb}}
\def\cRb{{\cal \Rb}}
\def\cWb{{\cal \Wb}}

\def\fd {{\rm N}}
\def\afd {{\overline{\rm N}}}

\def \II {I\hspace{-.1em}I\hspace{.1em}}
\def \IIA {\mbox{\II A\hspace{.2em}}}
\def \IIB {\mbox{\II B\hspace{.2em}}}
\def \gs {g^s}
\def \ls {\lambda^s}

\def \I {{\cal I}}
\def \qs {q\hspace{-.53em}/\hspace{.15em}}
\def \ks {k\hspace{-.53em}/\hspace{.15em}}
\def \YM {{\mbox{\tiny YM}}}
\def \gym {g_{\YM}}

\def \Lc {\L_c}
\def\IR{\relax{\rm I\kern-.18em R}}
\def \id {{\bf 1}}

\def\cci{\ell}
\def\ccj{\ell'}

\def\bbq{\pmb{q}}

\begin{document}
\begin{titlepage}
\begin{flushright}
\hfill{ NCTS-TH/1801} 
\end{flushright}
\hfill

 \begin{center}

{\Large \bf 
Soft Hair of Dynamical Black Hole and \\Hawking Radiation 
}\\[10mm]

{\bf Chong-Sun Chu${}^{1,2}$,  Yoji Koyama${}^1$}

{\itshape ${}^1$ Physics Division, National Center for Theoretical
  Sciences, \\
 National Tsing-Hua University, Hsinchu, 30013, Taiwan}\\[1mm]
{\itshape ${}^2$ Department of Physics, National Tsing-Hua
  University,  Hsinchu 30013, Taiwan}

{\small \sffamily
cschu@phys.nthu.edu.tw~, koyama811@cts.nthu.edu.tw
\\
}
\end{center}

\date{\today}

\begin{abstract}
Soft hair of black hole has been 
  proposed recently
  to play an important role
  in the resolution of the black hole information paradox. Recent work
  has emphasized that the soft modes cannot
  affect the black hole S-matrix due to
  Weinberg soft theorems.
  However as soft hair is generated by supertranslation
  of geometry which involves an angular dependent shift of time,
  it must have  non-trivial quantum effects.
  We consider supertranslation of the Vaidya black hole and
  construct a non-spherical
  symmetric dynamical spacetime with soft hair. We show that this spacetime
  admits a trapping horizon and is a dynamical black hole.
  We find that Hawking radiation
  is emitted from 
  the trapping horizon of the dynamical
  black hole. 
  The Hawking radiation 
  has a spectrum which depends on the soft hair of the black
  hole
  and this is consistent with the
  factorization property of the black hole S-matrix.
  
\end{abstract}

\end{titlepage}
\newpage


\section{Introduction}

Black hole model in general relativity identifies deep insufficiency
in our understanding of gravity.
As explained by Bekenstein \cite{Bekenstein:1973ur} and Hawking
\cite{Hawking:1974rv,Hawking:1974sw}, 
black holes obey the first law of thermodynamics \cite{Wald:1993nt},
with a temperature
that arises from the quantum process of Hawking radiation.
From the no-hair theorem
\cite{Chrusciel:2012jk},
one would expect
that the Hawking radiation to be completely independent of the state
of  matters entering the black hole. Therefore if the matters were in a
pure quantum state, and since the Hawking radiation is 
described by a mixed thermal state, the black hole
evaporation process would be non-unitary
which is in contradiction with quantum mechanics. 
See, for example,
\cite{Mathur:2009hf,Marolf:2017jkr} for a review of the black hole
information problem.

Recently, Strominger initiated a 
study of the infrared structure of gravity and
its connection with the asymptotic BMS symmetry
\cite{Strominger:2013jfa,  He:2014laa,Cachazo:2014fwa,
  Kapec:2014opa,   Strominger:2014pwa,  Strominger:2017zoo}.
Furthermore,
Hawking, Perry and Strominger
\cite{Hawking:2016msc,Hawking:2016sgy} and Strominger \cite{Strominger:2017aeh}
advocated 
a new approach to the black hole information paradox
based on 
a new kind of black hole soft hair.

BMS symmetry \cite{Bondi:1962px,Sachs:1962wk}
is the symmetry of asymptotically flat spacetime 
generated by supertranslation, 
an angular dependent shift of the time coordinate in the asymptotic
region.
With the BMS symmetry, asymptotically flat spacetimes are
characterized by, in
addition to the standard ADM charges of mass, electric charge and
angular momentum 
\cite{Arnowitt:1962hi}, also an infinite number of BMS charges of
supertranslation and superrotation charges 
\cite{Wald:1999wa,Barnich:2011mi}.
The existence of an infinite number of new charges for asymptotically
flat spacetime is a very interesting observation. When applied to
black hole in Minkowski spacetime, this means 
that supertranslated black
hole carries an infinite amount of soft hair
characterized by the BMS
charges.
The existence of soft hair is not a contradiction to the standard no-hair
theorem
of black hole since the supertranslated black hole is diffeomorphic to
the untranslated one, and so they carry the same ADM
charges.
Nevertheless soft hair modifies the
definition of time at null infinity and it acts  non-trivially on the
classical phase space. It is physical and its effects
can  be observed classically with the
gravitational memory effect
\cite{Zeldovich,Braginsky:1986ia,Braginskii:1987,Blanchet:1992br}.

Soft hair can be generated via physical process. 
In the case of Schwarzschild black hole, 
\cite{Hawking:2016sgy} showed
that one could grow soft hair on it
by throwing in a shockwave of radiation. This relation of the
soft hair with the collapsed matter suggests that  soft hair could
encode information about the  black hole creation and evaporation process.
It was further argued  \cite{Hawking:2016msc} that,
if viewed as a
scattering amplitude
in the quantum theory, the process of the black hole formation and
evaporation should be constrained by the infinite symmetries
of BMS and hence the existence of an infinite amount of
soft hair could help to resolve the information paradox.
The problem of how the black hole S-matrix
could be constrained by the
soft modes has been
further analyzed by a number of authors recently. 
As emphasized by \cite{Mirbabayi:2016axw,Gabai:2016kuf, Gomez:2016hxz,
  Bousso:2017dny},  the
scattering of the hard modes (like the black hole process)
is factorized from the scattering of the soft modes. The decoupling
can be seen in the dressed state approach. The decoupling has
also been argued to be a generic feature of  quantum gravity, at least
perturbatively \cite{Donnelly:2017jcd}.

Although
the S-matrix turns out not to be a good observable for
studying the evaporation process of black hole,
it does not exclude soft hair from having physical
effects on the black hole physics.
Quantum mechanically, the change of
time is translated to a change of the quantum vacuum, so soft hair is
expected to leave quantum effects on the black hole physics. 
The main  motivation of our work is to find out
how and
what aspects of the quantum physics of black hole is affected by its
soft hair.

In this regard, we consider the effects of soft hair on the spectrum
of the Hawking radiation.
For the soft-hairy Schwarzschild black hole, we find that
the Hawking radiation is insensitive to the
soft hair. In this process, soft hair is implanted on the eternal
Schwarzschild black hole with an energy flux of shock wave.
However this is not entirely consistent as Hawking radiation carries
away energy and should backreact on the metric. One should allow for
the time dependence of the metric and consider a
dynamical black hole. We therefore consider the more realistic process
of black hole evaporation due to  Hawking radiation
of this dynamical black hole and
investigate how the soft hair would affect the Hawking radiation.
The black hole evaporation process can be modeled with a supertranslated 
Vaidya spacetime.
We employ the tunneling method to compute the Hawking radiation and
find that it has a dependence on the soft hair configuration.
Our result is consistent with the factorization
property of the S-matrix: the final state of black hole
evaporation consists of Hawking radiation plus a sea of soft modes.
If one use the dressed states as a basis for the soft modes,
then the Hawking radiation would have a spectrum 
which is independent of the
soft hair configuration. 
On the other hand, if one use the undressed
states as in our computation, then the Hawking 
radiation would develop a dependence on the soft hair.

The organization of the paper is as follow. In section 2, we introduce
the Vaidya spacetime and its supertranslation. We show that this spacetime
can be obtained from the collapse
of a certain energy momentum tensor with soft charges. 
In section 3, we discuss some properties of the supertranslated Vaidya
black hole.  We briefly review the definition of trapping horizon for
dynamical black holes and show that the supertranslated Vaidya spacetime
is a dynamical black hole with a trapping horizon. 
Next we define the horizon surface gravity associated with the trapping horizon
of the supertranslated Vaidya black hole by extending the definition
of the horizon surface gravity of spherically symmetric black holes
using Kodama vector. In section 4, we apply the Hamilton-Jacobi
tunneling method \cite{Srinivasan:1998ty}, 
which has been used as frequently as the null geodesic method developed 
by Parikh and Wilczek \cite{Parikh:1999mf},
 to compute the Hawking radiation from 
the supertranslated Vaidya black hole. The Hawking radiation spectrum
takes the standard form and the dependence on soft hair is encoded in
the temperature of the radiation. We discuss the results with respect
to dressing of soft modes in the final state of observation.
We conclude the paper with some
further discussion.

\section{Supertranslation of Vaidya Spacetime}

\subsection{Supertranslated Vaidya spacetime}


Let us start with a brief review on supertranslation
for an asymptotically flat metric in four dimensions. 
At the infinity, depending on the
physical situation one wants to describe, one may impose
different falloff conditions on the metric. 
In general one wants to
choose the falloff conditions such that interesting solutions such as
gravitational radiations are included, but unphysical solutions
(e.g. those with infinite energy) are ruled out. The choice of falloff 
conditions of Bondi,
van der Burg, Metzner and Sachs (BMS)
\cite{Bondi:1962px,Sachs:1962wk,Barnich:2009se,Sachs:1962zza}
considers metric with
the asymptotic expansion near the past null infinity ${\cal I}^{-}$
\cite{Hawking:2016sgy},
\bea \label{bmetric-}
ds^2 &=& - dv^2 + 2 dv dr + r^2 \gamma_{AB} d \Th^A d\Th^B \nn\\
& &+ \frac{2m}{r} dv^2 + r C_{AB} d \Th^A d\Th^B
+\frac{1}{4} \gamma_{AB} C_{CD} C^{CD} d \Th^A d \Th^B \nn \\
& &- D^B C_{AB} dv d \Th^A -\frac{1}{r}\left(
\frac{4}{3}N_A -\frac{4v}{3}\del_A m - \frac{1}{8}\del_A(C_{BD}
C^{BD})\right) dv d\Th^A \nn \\
& & -\frac{1}{16r^2} C_{AB} C^{AB} dv dr, 
\eea
where $\gamma_{AB}$ is the metric on the unit two sphere
and $\Theta^A = (z,\zb)$ are the angular coordinates.
%
In \eqref{bmetric-} the Bondi mass aspect $m$, the traceless tensor $C_{AB}$
and the angular momentum aspect $N_{A}$ depend on $(v,\Theta^A)$.
BMS supertranslation is the diffeomorphism which preserves the Bondi
gauge
and the asymptotic falloff conditions. It is generated by
the vector field: 
\be
\zeta_{f}=f\partial_{v}-\frac12 D^{2} f\partial_{r}+\frac{1}{r}D^{A}f\partial_{A},
\label{zeta}
\ee 
where $D_{A}$ is the covariant derivative with respect to
$\gamma_{AB}$. 
%
%
Supertranslations are 
characterized by an arbitrary function of the angular
variables, $f=f(\Theta)$. 

We are interested in dynamical black holes in asymptotically flat spacetime.
An example is the Vaidya metric.
There are two kinds of Vaidya spacetime, one written in terms of the
retarded (outgoing) null coordinates  and one in terms of the advanced
(ingoing) null coordinates.
The Vaidya metric in the advanced Bondi
coordinates $(v,r,\Theta^{A})$ is given by
\be
ds^{2}=\gb_{\mu\nu}dx^{\mu}dx^{\nu}
=-Vdv^{2}+2dvdr+r^{2}\gamma_{AB}d\Theta^{A}d\Theta^{B},
\qquad
V\equiv 1-\frac{2M(v)}{r}.
\label{vaidya}
\ee
The Bondi mass aspect $M = M(v)$ is a function of the advanced time $v$.
The ingoing Vaidya metric satisfies the Einstein equation with the
energy flux (for Newton constant $G_N =1$)
\be
{\bar T}_{vv}=\frac{M'(v)}{4\pi r^{2}}, \qquad M':=\partial_{v}M(v).
\ee
Null energy condition implies that $M'(v) \geq 0$ which
corresponds to ingoing energy flux being absorbed by the black hole. This
geometry naturally
describes the formation of a black hole by the collapse of matter.
However with the sign reversed, $M' <0$, the metric can also be taken
as a model describing the evaporation of black hole by an outgoing
energy flux. In this paper,  we will determine the Hawking radiation
from the Vaidya black hole with soft hair
using the tunneling method. We note that in order to
utilize the tunneling method, one must employ  a metric which is
smooth across the
location where tunneling occurs, i.e. the horizon. This is
suitable for the ingoing Vaidya metric which covers the both the
interior and exterior of the black hole; but not for the outgoing Vaidya
metric since  it covers the exterior region of a black hole and the
interior region of a white hole. Therefore we will employ
the ingoing Vaidya metric with $M' <0$ as a model to discuss the
evaporation of dynamical black hole due to Hawking radiation
\footnote{
Most part of our analysis  actually holds true for any sign of
$M'$. For example, with $M'>0$, the results, \eq{kappa}, \eq{emsv} give
the influence of the soft hair on the Hawking radiation
during the stage the black hole was formed. However the identification
\eq{mloss} of the
energy loss from Hawking radiation with the change of mass of the
black hole 
holds only in the case of mass loss $M'<0$. }.


The supertranslated Vaidya metric 
$g_{\mu\nu}$
is  obtained by acting the vector field $\zeta_{f}$ of \eqref{zeta} on
the Vaidya metric \eqref{vaidya},
\be \label{gtg}
g_{\mu\nu}=\gb_{\mu\nu}+{\cal L}_{\zeta_{f}}\gb_{\mu\nu}.
\ee
The result is
\begin{align}
g_{\mu\nu}dx^{\mu}dx^{\nu}
&=
-\Big(V-\frac{2fM'}{r}-\frac{MD^{2}f}{r^{2}}\Big)dv^{2}
+2dvdr
-D_{A}(2Vf+D^{2}f)dvd\Theta^{A}
\notag
\\
 &
 \quad
+(r^{2}\gamma_{AB}+2rD_{A}D_{B}f-r\gamma_{AB}D^{2}f)d\Theta^{A}d\Theta^{B}.
\label{svaidya}
\end{align}
Note that the metric \eq{svaidya} can be
extended to
finite distance and is a solution to the linearized
Einstein equations for all $r$. 
In this paper, 
we restrict ourselves to linearized theory in the metric perturbation,
or equivalently in $f$,
which
means that the back reaction
caused by the energy momentum tensor of the
gravitational waves and the quantization of
gravitational fluctuations will not be considered in our analysis.
From the form of \eqref{svaidya},
it is clear that the supertranslated Vaidya spacetime is
non-spherical.
This may also be seen physically from the fact that
the metric \eq{svaidya} can be obtained, as we will show in the
next subsection, by throwing in a
non-spherically symmetric energy
momentum flux to the Vaidya spacetime.
We  also remark that unlike
the static case of the supertranslated Schwarzschild spacetime, 
the supertranslated Vaidya spacetime does not have an
event
horizon. In fact for a dynamical black hole, the concept of
event
horizon has
to be generalized.

In the next section, we
will review the definition of dynamical black holes  
and show that the
supertranslated Vaidya spacetime admits a trapping horizon and
describes a non-spherical dynamical black hole. As far as we are aware of,
\eq{svaidya} is the first example of a non-spherical dynamical
black hole. Before that, let us first demonstrate that the
supertranslated hair configuration \eq{svaidya} can be obtained by throwing
in a shockwave-like
energy momentum flux to the Vaidya black hole.

\subsection{Implantation  of supertranslation hair}

The spacetime \eq{svaidya} describes a black hole 
with supertranslation hair implanted on it. Physically, the configuration of
soft hair can be implanted by throwing in an energy flux of a particular form.
Consider 
an ansatz for the perturbed energy momentum tensor,
\begin{align}
{\hat T}_{vv}&
=\frac{1}{4\pi r^{2}}\left(\hat{\mu}
(v,\Theta)+{\hat T}(\Theta)\delta(v-v_{0})\right)
+\frac{1}{4\pi r^{3}}\left({\hat T}^{{(1)}}(\Theta)\delta(v-v_{0})
+{\hat t}^{(1)}(\Theta)\theta(v-v_{0})\right),
\notag\\
{\hat T}_{vA}&
=
\frac{1}{4\pi r^{2}}
\left({\hat T}_{A}(\Theta)\delta(v-v_{0})
+{\hat t}_{A}(\Theta)\theta(v-v_{0})\right),
\label{perem}
\end{align}
where $\theta(v-v_{0})$ is step function.
The covariant conservation ${\bar \nabla}^{\mu}{\hat T}_{\mu\nu}=0$
imposes that
\be
D^{A}{\hat T}_{A}(\Theta)={\hat T}^{{(1)}}(\Theta),\quad
D^{A}{\hat t}_{A}(\Theta)={\hat t}^{{(1)}}(\Theta),
\label{conservation}
\ee
where ${\bar \nabla}_{\mu}$ stands for the covariant derivative
with respect to the background Vaidya metric $\gb_{\mu\nu}$.
In addition, we consider energy momentum tensor satisfying the relation
\be
(D^{2}+2){\hat T}^{{(1)}}(\Theta)=-6M(v){\hat T}(\Theta).
\label{relation}
\ee

Now for general asymptotically flat metric 
\eq{bmetric-}
in the Bondi gauge, the Einstein equations at ${\cal I}^{-}$ give the following
constraints at order ${\cal O}(r^{-2})$:
\begin{align}
\partial_{v}m
&=  \frac{1}{4}D^{A}D^{B}\partial_{v}C_{AB}
+4\pi \lim_{{r\to \infty}}(r^{2}T_{vv})
,
\label{evv}
\\
\partial_{v}N_{A}
&=
v D_{A}\partial_{v}m+
\frac14 D^{B}(D_{B}D^{C}C_{AC}-D_{A}D^{C}C_{BC})
-{8\pi \lim_{r\to\infty} (r^{2} T_{vA})},
\label{eva}
\end{align}
where we neglected quadratic perturbation terms  off from flat space.
Our claim is that one can reproduce the supertranslated Vaidya metric
with an appropriate choice of the energy momentum tensor
\be
T_{\mu\nu}
=
{\bar T}_{\mu\nu}
+
{\hat T}_{\mu\nu}.
\label{embp}
\ee
To start, the constraint \eq{evv} can be solved by taking
\be
\frac{1}{4}D^{A}D^{B}\partial_{v}C_{AB}
= - {\hat T}(\Theta)\delta(v-v_{0})
\label{hatT}
\ee
and
\be
\hat{\mu}(v,\Theta) =
\partial_{v}\left(\theta(v-v_{0})
({\hat C}M'+{ \mu})\right),
\ee
where ${\mu}$ is a constant and ${\hat C}={\hat C}(\Theta)$ is an
arbitrary function of angles. 
Then
\be
m  = M + \theta(v-v_{0})({\hat C} M' +{\mu} ).
\ee
As for the equation \eq{hatT}, let us consider
\be
   {\hat T}(\Theta)
=-\frac{1}{4}D^{2}(D^{2}+2){\hat C},
\ee
which implies, from 
\eqref{conservation} and \eqref{relation}, that
\be
{\hat T}^{(1)}(\Theta)
=\frac{3M}{2} D^{2}{\hat C},
\qquad
{\hat T}_{A}(\Theta)
=\frac{3M}{2} D_{A}{\hat C}.
\ee
Then \eq{hatT} can be solved and gives
\be
C_{AB} 
=
\theta(v-v_{0})
(2D_{A}D_{B}{\hat C} - \gamma_{AB}D^{2}{\hat C}).
\label{wanted}
\ee
In turn from \eq{eva},  we obtain
\be
N_{A}
=
\theta(v-v_{0})(vM' -3M) D_{A}{\hat C},
\label{wanted2}
\ee
and
\be
   {\hat t}_{A}(\Theta)= M' D_{A}{\hat C},
   \qquad {\hat t}^{(1)}(\Theta)=M' D^{2}{\hat C}.
\ee
where ${\hat t}^{(1)}$ is obtained by solving \eqref{conservation}.

The energy momentum tensor ${\hat T}_{\mu\nu}$ is shockwave-like, with
a delta function and a step function component.  
And the corresponding spacetime $g_{\m\n}$ can be written as a perturbation over
the background Vaidya metric of mass $M(v)$:
\be \label{g-pert-}
g_{\m\n} = \gb_{\m\n} + h_{\m\n}
\ee
with the perturbations:
 \begin{align} 
 h_{vv}
 &=\theta(v-v_{0})
 \left(\frac{2}{r}({\mu}+\Ch M')+\frac{1}{r^{2}}MD^{2}\Ch\right),\notag\\
 h_{vA}&=-\theta(v-v_{0})D_{A}
 \left(V \Ch+\frac12 D^{2} \Ch\right),\notag\\
 h_{AB}&=r\theta(v-v_{0})
 \left(2D_{A}D_{B}\Ch-\gamma_{AB}D^{2}\Ch\right).
 \label{spert}
 \end{align}
 Like in the static case \cite{Hawking:2016sgy}, the perturbation can be written as
 \be 
 h_{\m\n} = \th(v-v_0) \left(
\cL_ {f=\Ch} \gb_{\m\n} + \frac{2\mu}{r} \d_\m^v \d_\n^v
 \right).
 \ee
 Therefore our shockwave-like energy momentum flux creates a
 supertranslation of the Vaidya
 metric with the supertranslation parameter $f =\Ch$. Also it 
 shifts the mass parameter by a
 constant amount in case $\mu$ is nonzero.
 Physically it means that during the process of the formation of the
 dynamical Vaidya black hole,
 we can also implant  a configuration of soft hair on it with the aid
 of the energy momentum flux 
 ${\hat T}_{\m\n}$.
 Note that
 the metric \eq{g-pert-} actually
 satisfies the linearized Einstein equations for all $r$ 
 and thus describes linear perturbations not only at around ${\cal
   I}^{-}$ but also in the interior of spacetime.
%

 The fact that the metric  \eq{g-pert-}  holds for all $r$,
 especially in the neighborhood of the black hole trapping horizon,
 has important physical consequences. 
 Below we will extend
 the consideration of Hawking radiation for dynamical black hole and show that
 Hawking radiation are created at the dynamical horizon of the soft hair
 implanted Vaidya black hole. Naturally the Hawking radiation spectrum can be
 expected to have
 dependence on the soft charges
 (superrotation charges) of the black hole. 
 We will show that this is indeed the case.

We end this section by noting that
while in this paper we focus on BMS supertranslations as asymptotic symmetries
defined at null infinity and its effect on the physics at the black hole horizon, 
there is a different sort of supertranslations (and superrotations) 
as asymptotic symmetries defined at black hole horizons, 
also known as horizon supertranslations, 
which has been studied for example in 
\cite{Hawking:2016msc,Koga:2001vq,Hotta:2000gx,Donnay:2015abr,Averin:2016ybl,Eling:2016xlx,Donnay:2016ejv,Cai:2016idg,Afshar:2016kjj}. 
Recent work on gravitational memory and quantum mechanical effects 
associated with the horizon supertranslations on Rindler horizon
can be found in \cite{Hotta:2016qtv,Kolekar:2017yoi,Kolekar:2017tge}.

\section{Properties of supertranslated Vaidya black hole} \label{sec3}

In this section we investigate some properties of the supertranslated 
Vaidya black hole
which will be helpful for interpreting and understanding of 
the results of Hawking radiation 
from the supertranslated 
Vaidya black hole obtained in the  section \ref{sec4}.

\subsection{Trapping Horizon}

The event horizon of a spacetime is a global concept and requires a knowledge
of the entire causal structure of spacetime. This is physically impossible
unless the spacetime is stationary and nothing changes. In this case one can
invoke the theorem of Hawking which says that the event horizon of a stationary
asymptotically flat spacetime is a Killing horizon. In addition, using the
null Killing vector,  one can associate
with the Killing horizon a surface gravity which plays the role of
temperature in the black hole thermodynamics and Hawking radiation effect.

However the situation becomes much more complicated in
the general dynamical case where
Killing horizon generally does not exist.
Much efforts have been spent on
looking for appropriate local definitions of horizon for dynamical
black hole spacetime
\cite{Hayward:1993wb,Ashtekar:1998sp,Ashtekar:2002ag,Booth:2003ji} 
(see also
\cite{Ashtekar:2004cn,Booth:2005qc,Hayward:2008ti,Gourgoulhon:2008pu,
  Faraoni:2013aba} for reviews).
These local horizons are typically defined in terms of trapped surfaces,
which are space-like 2-surfaces
for which the expansion of the outgoing null rays normal to the surface
vanishes. Let us review the precise definition here.
Consider a bundle of null geodesics with tangent vector 
$l_{-}^{\mu} $ and satisfies $g_{\mu\nu}l_{-}^{\mu}l_{-}^{\nu}=0$,
$l_{-}^{\mu}\nabla_{\mu}l_{-}^{\nu}=0$. Let us pick another null vector field
$l_+^{\mu} $ such that $g_{\m\n} l_{+}^\mu l_{+}^{\nu}=0$ 
and a relative normalization $l_{-}{}^{\mu}l_{+ \mu}=-1$. Then the metric in the
two-space $S$ orthogonal to $l_+$ and $l_-$ can be written as
\be
q_{\mu\nu}
=
g_{\mu\nu}+l_{- \mu}l_{+ \nu}+l_{- \nu}l_{+ \mu}.
\label{q}
\ee
By construction, $l_\pm$ is orthogonal to $S$:
$q_{\mu\nu}l_{\pm}^{\nu}=0$.
In general given a null flow characterized by the tangent vector $l_\mu$, one
can characterize the flow with a shear, rotation and an expansion
part. Of particular interests to general relativity is the
expansion scalar which is defined by the divergence of the flow: 
\be
\theta := q_{\m\n} \nabla^\m l^\n .
\ee
Physically, 
$\theta$ measures the expansion rate of the infinitesimally nearby
surrounding radial null geodesics \cite{Wald:1984rg}: the
bundle of null geodesics is expanding
if $\theta>0$ (gravity is not so strong) and contracting if $\theta<0$.
Now a space-like closed and orientable two surface in four-dimensions
has two independent normal
directions, corresponding to the ingoing and outgoing null rays. It is
thus natural to take $l_{-}^{\mu}$ and
$l_{+}^{\mu}$ to be the  
tangent vectors of the bundles of ingoing 
and outgoing null geodesics, and use their behaviour to characterize
the gravitational field surrounding $S$. A normal surface would have
$\th_- <0$ and $\th_+ >0$. A {\it trapped surface} is one
with $\th_-<0$ and $\th_+ < 0$, i.e. the outgoing null rays are
contracting at $S$ instead of expanding. The surface is {\it
  marginally trapped } if $\th_- < 0$ and $\th_+ = 0$,
i.e. the outgoing null rays
momentarily stop expanding. 
Trapped surfaces are interesting since
under certain physically reasonable assumptions, 
they lead to the presence of singularities. The cosmic
censorship hypothesis then suggests that there must be an event
horizon, with the trapped surface located inside 
of certain black hole
horizon. 
This is a highly non-trivial problem in the dynamical case
since when embedded in a dynamical spacetime,
the horizon is not
expected
to be a null hypersurface although it should still exhibit infinite
redshift. Various definitions of black hole
horizon have been proposed and considered. Among them of particular
importance is the future outer trapping horizon (FOTH) introduced by Hayward
\cite{Hayward:1993wb}, where later refinements and generalizations are based on. 
In this paper, we find
that the superstranslated Vaidya spacetime is a dynamical black hole
with a  future outer trapping horizon of Hayward.

A future outer (marginally) trapping horizon 
is a smooth
three-dimensional submanifold of spacetime which is foliated by closed
space-like surfaces $S_t$, $t\in \bR$, with null normals $l_\pm$
constructed by ingoing and outgoing rays such that:
\be
\theta_{+}=0 \quad (\text{marginally trapped}),
\quad
\theta_{-}<0 \quad (\text{future type}),
\quad
l_-^{\mu}\del_{\mu}\theta_{+}<0 \quad (\text{outer type}).
\label{mts}
\ee
The first condition in \eqref{mts} 
specifies the location of 
marginally trapped surfaces where nearby surrounding radial 
outgoing null geodesics are parallel.
The second condition says that the trapping horizon is of future type, i.e.
a black hole rather than a  white hole.
The third condition says a motion of $S_t$ along $l_-$ makes it
trapped, hence it is outer rather than inner kind.

Now let us apply these concepts to 
the supertranslated Vaidya spacetime. Let us consider
radial ingoing and outgoing null geodesics. 
The corresponding tangent vectors are
\be
l_{-}^{\mu}=(0,-1,0,0),\qquad
l_{+}^{\mu}=\left(1,\frac12\Big(V-\frac{2fM'}{r}
-\frac{MD^{2}f}{r^{2}}\Big),0,0 \right).
\label{lm}
\ee 
Using \eqref{svaidya}, \eqref{q} and \eqref{lm}, we get
\be
\theta_{-}=-\frac2r, \qquad
\theta_{+}=\frac1r \Big(V-\frac{2fM'}{r}-\frac{MD^{2}f}{r^{2}}\Big),
\label{theta}
\ee
where we have discarded terms of ${\cal O}(f^{2})$.
For the supertranslated Vaidya black hole,
the first condition in \eqref{mts} 
yields 
\footnote
{Actually $\theta_{+}=0$ has two solutions. One is given by \eqref{rh} and
 corresponds to a marginally trapped surface of outer type as discussed above.
The other solution $r=-\frac12 D^{2}f$  
corresponds to a marginally trapped surface
of inner type with $l^{\mu}\partial_{\mu}\theta_{+}>0$,
whose foliation gives a future inner trapping horizon
if $r=-\frac12 D^{2}f>0$.}
\be
r=r_{h}=2M+2fM'+\frac12 D^{2}f.
\label{rh}
\ee
In our calculation, $f$ is taken as a small perturbation and
the $f$ dependent terms give correction to the Schwarzschild radius.
For consistency, we will consider the case where the corrections are small 
such that the first term dominates, $M \gg | M' f|, |D^2 f|$,
and  so $r_h >0$.
The second condition in \eqref{mts} is trivially satisfied.
The third condition yields 
\be
-\frac{1}{2 r_{h} M}\Big(1-f\frac{M'}{M}\Big)<0,
\label{outer}
\ee
which is satisfied for the same assumption as above. 
The condition \eqref{outer} is actually
equivalent to a positive horizon surface gravity
which we will discuss in the next section. 
Therefore, a trapping horizon foliated by marginally trapped surfaces
$\eqref{rh}$ is of future and outer type
and is a FOTH of the supertranslated Vadiya black hole. 
A FOTH is in general not null but it still has infinite red shift. 
In fact,
substituting \eqref{rh} into \eqref{svaidya} and fixing angular coordinates
 $d\Theta^{A}=0$ 
 gives
\be
ds^{2}=
4\left(M'+fM''+\frac14 M' D^{2}f\right)dv^{2}.
\ee
Hence the FOTH of supertranslated Vaidya black hole
at a fixed angular coordinates is time-like if 
$M'<-fM''-\frac14 M' D^{2}f$ for which the null energy condition will be
violated and space-like if $M'>-fM''-\frac14 M' D^{2}f$.
For the stage of evaporation the black hole mass will be decreasing $M'<0$ 
due to ingoing negative energy flux and the FOTH would be time-like.
%

\subsection{Surface Gravity}

Next we discuss the  horizon surface gravity 
associated with the trapping horizon of the supertranslated Vaidya
black hole. 

For stationary black hole, surface gravity is defined
in terms of the null Killing vector of the Killing horizon.
In a time-dependent spacetime, there is generally no asymptotically time-like
Killing vector to define a preferred time coordinate. As a result, quantities
such as four-acceleration and surface gravity cannot be defined unambiguously.
Instead,
depending on the local definition of horizons that is adopted,
one may define a notion of surface gravity correspondingly,
see e.g. 
\cite{Faraoni:2013aba,Nielsen:2007ac,Vanzo:2011wq}. 
For the FOTH of Hayward, there is
a quite explicit and satisfactory definition of surface gravity
for the spherically symmetric case. In fact, for any 
spherically symmetric metric, 
one can show the existence of
a unique vector field $K^\m$,
called the Kodama vector, which satisfies 
\cite{Kodama:1979vn}
\be
\nabla^{\nu}(G_{\mu\nu}K^{\mu})=0, \qquad \nabla_{\mu}K^{\mu}=0.
\label{kodama}
\ee
Here $G_{\mu\nu}$ is the Einstein tensor. 
Kodama vector gives a preferred time direction for dynamical spacetimes
in spherical symmetry.
In an asymptotically flat spacetime, with an appropriate normalization,
the Kodama vector coincides with the time translation Killing vector at spatial infinity. 
Hayward has shown that the following relation holds for
Kodama vector \cite{Hayward:1997jp}, 
\be
K^{\mu}\nabla_{[\nu}K_{\mu]}= -\kappa K_{\nu}
\label{sgr}
\ee
where the equality holds when it is evaluated on a trapping horizon.
The coefficient $\kappa$ defines the horizon surface gravity of the FOTH
of Hayward.
Kodama vector coincides with the time translation Killing vector field 
for stationary solutions, and 
the defining equation for surface gravity \eqref{sgr} 
reduces to that of Killing surface gravity
in static limit by virtue of Killing equation and 
thus can be used to define surface gravity $\kappa$
for dynamical black holes with spherical symmetry.
For non-spherical dynamical black holes, it is not known in general whether the
Kodama vector exists. However if it exists, we can use \eq{sgr}
to define the surface gravity.

For a spherically symmetric Vaidya spacetime, 
the Kodama vector is ${\bar K}^{\mu}=\delta^{\mu}_{v}$ 
and the horizon surface gravity of Vaidya black hole 
defined by \eqref{sgr} is 
${\bar \kappa} =\frac{1}{4M(v)}$. 
We would like to extend the definition of surface gravity
to the supertranslated Vaidya black hole case using the Kodama vector.
For the supertranslated Vaidya metric $g_{\m\n}$
\eq{svaidya}, although it is not
spherically symmetric, we find that
\be
K^{\mu}
=\delta^{\mu}_{v},
\qquad
K_{\mu}=g_{\mu\nu}K^{\nu},
\label{skodama}
\ee
solves the equation \eqref{kodama}
up to ${\cal O}(f^{2})$ term. 
Evaluated on the FOTH,
the surface gravity in \eqref{sgr}
is found to be
\be
\kappa 
= \frac{1}{4M}\Big(1-f\frac{M'}{M}\Big),
\label{sgr2}
\ee
at the linear order in $f$.
Note that the Kodama-like vector 
\eqref{skodama}
coincides with 
the time-translation Killing vector for the supertranslated Schwarzschild 
in static limit.
In the supertranslated case, the horizon surface gravity of the
supertranslated Vaidya black hole 
has a dependence on the supertranslation hair and is
not a constant on the FOTH. Physically this means 
that
the 
black hole does not correspond to a system in thermal equilibrium.
In general one cannot define
a specific temperature for 
a nonequilibrium system.
Nevertheless
one may still define a temperature locally provided that the system
is in equilibrium locally. We propose to interpret $\k$ as
a local measure of the 
temperature of the superstranslated Vaidya black hole.
As we mentioned above, the outer trapping horizon
corresponds to the positive surface gravity $\kappa$ given by
\eqref{sgr2}. 
The inner trapping horizon would correspond to $\kappa<0$ and
the degenerate (extremal) one corresponds to $\kappa=0$. 
It is interesting to think about what these thermodynamical notions mean in
terms of spacetime physics.


\section{Hawking Radiation 
from Supertranslated Vaidya Black Hole}  
\label{sec4}

\subsection{Hamilton-Jacobi Method}


Now let us consider Hawking radiation from the  supertranslated
Vaidya black hole. As argued by Parikh
and Wilczek \cite{Parikh:1999mf}, the
Hawking radiation can be computed as a tunneling process based on
null geodesic motion of particle in the black hole geometry.
That this is possible
without using the full-fledged
quantum field theory is because of the huge  red shift factor at the horizon of
the black hole. As a result, Hawking radiation 
observed at asymptotic infinity arises from emitted wave with
vanishing wavelength near the horizon, and therefore, as far as the
tunneling process is of concern, a
point particle approximation near the horizon is good.
We refer the readers to \cite{Vanzo:2011wq} for a review on the tunneling
methods and Hawking radiation. In addition to the original method of
null geodesic, the Hamilton-Jacobi method has also been developed to
compute the Hawking radiation of black hole.
It has been shown for stationary black hole spacetimes that  the two methods
provide the same result for 
the semi-classical emission rate at the leading order of the energy of 
the emitted particles \cite{Kerner:2006vu}.
While the null geodesic method is based on the self-gravitation (backreaction) 
of emitted particles and the energy conservation of the whole spacetime, 
the Hamilton-Jacobi method would be simple to compute the imaginary part 
of the particle action without reference to the self-gravitation of emitted 
particles. Also covariance of the method is well understood
in the Hamilton-Jacobi equation.
Moreover, the Hamilton-Jacobi method can be applied to either
stationary or dynamical black holes 
\cite{Visser:2001kq,DiCriscienzo:2007pcr,Hayward:2008jq}
so that it is suitable for our purpose.
We will therefore adopt the Hamilton-Jacobi method
to compute the Hawking radiation spectrum for the supertranslated Vaidya
black hole.

Consider now, for example, a
minimally coupled massless scalar particles, $\Box \phi(x)=0$. 
We look for a solution to this equation by a WKB ansatz 
$\phi = A(x)e^{iS/{\hbar}}+{\cal O}(\hbar)$.
At the leading non-trivial order of $\hbar$,
the Klein-Gordon (KG)
equation reduces to the Hamilton-Jacobi equation
\be
g^{{\mu\nu}}\partial_{\mu}S\partial_{\nu}S=0.
\label{hjeq}
\ee
Generally there are two solutions to this equation and they
correspond to the two different solutions of the KG
equation. 
Physically, they describe motion of a particle getting out
(outgoing solution) and falling in to the black hole (ingoing solution).
The motion of particle is given by a null geodesic in the black hole
background. 
One can then reconstruct the particle action by a line integral 
\be
S=\int_{P} dx^{\mu}\partial_{\mu}S,
\label{S}
\ee
where $P$ is the null geodesic.
$S$ is real in general, but can become complex if the trajectory is a
classically forbidden one. Since particles can fall into a black hole
along a classically permitted path, we expect $S$ to possibly become
imaginary only for the outgoing solution. 
In the case of static black holes, this occurs for the outgoing
solution along
a  trajectory crossing the event horizon 
which we call a tunneling path, and the imaginary contribution
arises from the residue of the  pole in $\partial_{\mu}S$
at the location of horizon.
To leading order in $\hbar$,
the semiclassical emission rate $\Gamma_{{\rm em}}$ is given by the WKB formula
\be
\Gamma_{{\rm em}} \propto \exp\Big(-2\,{\rm Im}S_{out}\Big),
\label{emission}
\ee
where $S_{out}$ denotes the particle action for the
outgoing solution and we have set $\hbar=1$.
For static black hole spacetimes, it is easy to show
that the semiclassical emission rate satisfies the
detailed balance relation 
\cite{Srinivasan:1998ty,Hartle:1976tp}
$\Gamma_{{\rm em}}=e^{-\beta \omega}\Gamma_{{\rm ab}}$  
where $\Gamma_{{\rm ab}}$ is the semiclassical absorption rate and
 $\omega$ is the energy of the emitted particles. This allows one
to have a thermal interpretation of the result with $\beta^{{-1}}$
being the (Hawking) temperature of the system.

Let us consider the semiclassical emission rate 
for supertranslated Schwarzschild black hole. In the
advanced Bondi
coordinates, the metric reads 
\cite{Hawking:2016sgy}
\begin{align}
ds^{2}=
& -\Big( V - \frac{M D^2 f}{r^2} \Big)dv^{2} + 2dv dr
-dv d \Theta^A D_A (2 V f + D^2 f) \notag\\
& +(r^{2}\gamma_{AB}+2rD_{A}D_{B}f-r\gamma_{AB}D^{2}f) d\Theta^{A}d\Theta^{B},
\end{align}
where $V =1-\frac{2M}{r}$.
This coordinate system has a nice feature for tunneling computation since
it is non-singular at the event horizon $r=2M+ \frac{1}{2} D^2 f$. As
shown by \cite{Hawking:2016sgy},
the metric can be extended to the interior and is
still a solution of the Einstein equations. 
We can see that this coordinate system covers both interior and 
exterior regions of a black hole and 
hence it is an appropriate coordinate system 
to compute the semiclassical emission rate for black holes. 
On the other hand, retarded 
Bondi coordinates would be
suitable for the computation of the semiclassical absorption rate for
white holes.
We consider radial null geodesics in $(v-r)$ plane 
for which the Hamilton-Jacobi equation is
\be
\Big( V - \frac{M D^2 f}{r^2}  \Big)(\partial_{r}S)^{2}
-2\omega\partial_{r}S
=0,
\label{swhj}
\ee
where
\be
\omega=-\xi^{\mu}p_{\mu}=-\xi^{\mu}\partial_{\mu}S=-\partial_{v}S.
\ee
$\xi$ is the time-translation Killing vector field $\xi=\partial_{v}$
and $p_{\mu}$ is the 4-momentum of particle. $\o$
is the energy of particle and is a constant in motion.
The solutions to \eqref{swhj} are
\be
\partial_{r}S_{{out}}=\frac{2\omega}{ V - \frac{M D^2 f}{r^2} },
\qquad
\partial_{r}S_{{in}}=0,
\ee
where $S_{out}$ corresponds to the outgoing solution ($\partial_{r} S>0$)
for $r>2M$, and
$S_{{in}}$ to the ingoing solution.
Using \eqref{S}, the corresponding action is
\be \label{sout}
S_{out}= -\int \omega dv + \int \frac{2\omega dr}{ V - \frac{M D^2 f}{r^2} },
\qquad
S_{{in}}= -\int \omega dv.
\ee 
As mentioned above, the ingoing action does not have an imaginary part
since  absorption is complete for black holes
in the classical limit \cite{Chatterjee:2007hc}. The integrand
for $S_{out}$
has 
poles at, up to order $f$ terms,
$ r= r_h := 2M + \frac{1}{2} D^2 f$ which is the location of
the (shifted) event horizon, and at $r= r_f :=  - \frac{1}{2} D^2 f$.

\begin{figure}[t!] 
\centering
\vspace{-1.5cm}
\includegraphics[width=5cm]{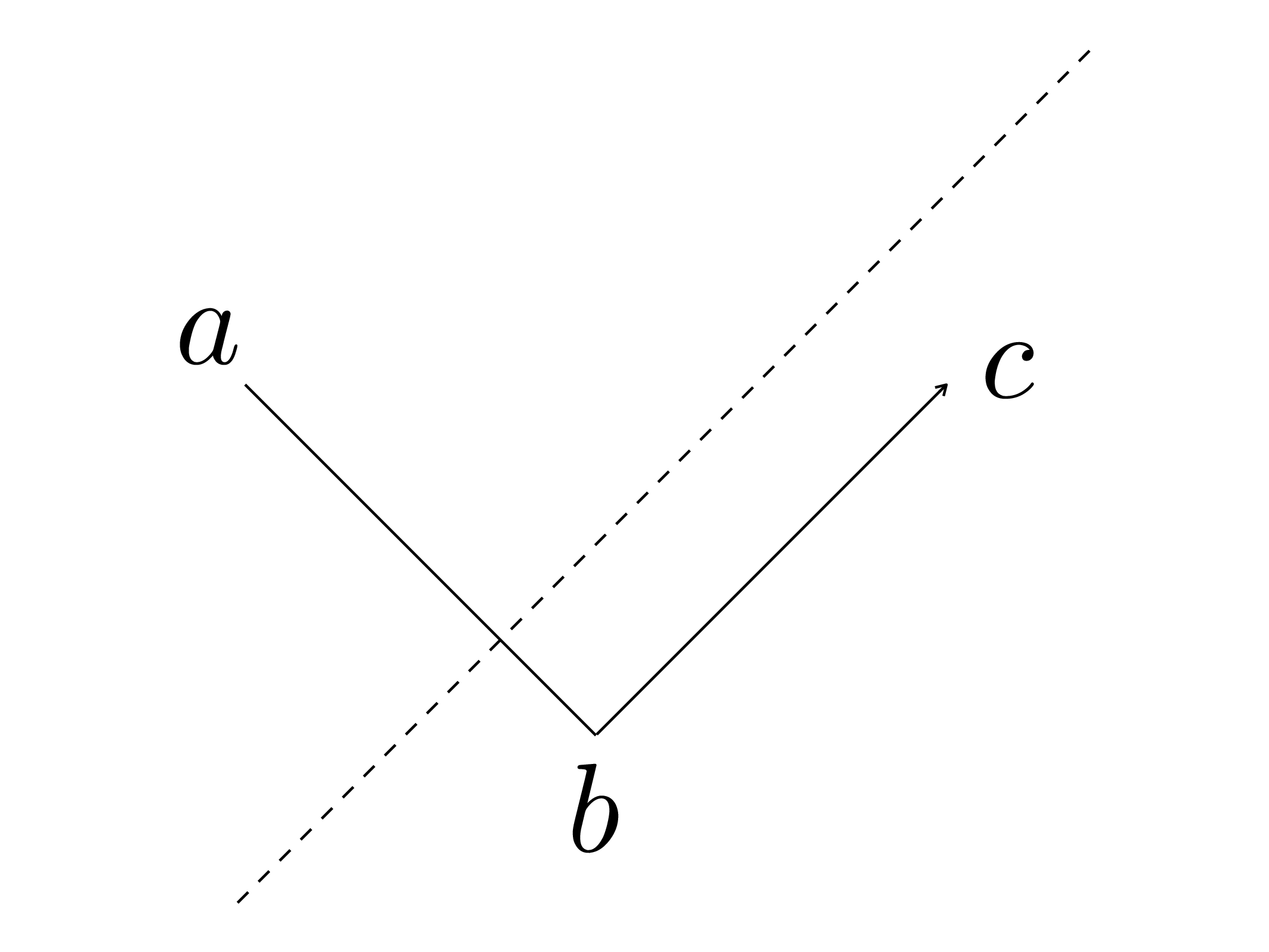}
\caption{A trajectory from inside to outside of the black hole.
The dashed line represents the event horizon.}
\label{fig:1}
\end{figure}

To obtain $S_{out}$, we need to perform the integration for along a
trajectory of motion.
As we consider s-wave motion, 
there are two types of path:
\bea
dv = 0, && \qquad \mbox{(type-I)} \nn\\
\frac{dr}{dv} = -\frac{1}{2} 
\left(V-\frac{MD^{2}f}{r^{2}}\right),  && \qquad \mbox{(type-II)}
\eea
In  Fig.\ref{fig:1}, we show a trajectory
$\overrightarrow{abc}$ 
which contains a  path $\overrightarrow{ab}$ 
moving backward in time and crossing the horizon. This part is
described by a type-I path.
After tunneling out from the horizon, the emitted particle escapes to
infinity on a type-II path
and would be observed as Hawking radiation \footnote{
This tunneling process can also be interpreted in a different way. 
A pair of virtual particles is created near $b$. 
One with negative energy falls into the black hole 
and the other with positive energy escapes to infinity \cite{Hartle:1976tp}.}.
It is easy to see that $S_{out}$ vanishes on a type-II path. For a
type-I path, we have
\be \label{S-out}
S_{out}=
\int_{\rm I}
\frac{2\omega r^{2} dr}
{(r-r_{h})(r-r_{f})},
\ee
As our path crosses the horizon $r=r_h$, the integral is divergent. 
Adopting a Feynman's $i\epsilon$
prescription to deform the integral
\footnote{
Here $i\epsilon$ is introduced in such a way that 
the positive energy exponential function decays and 
the relation
$\Gamma_{{\rm em}}=e^{-8\pi \omega M}\Gamma_{{\rm ab}}$
for black hole semiclassical emission rate is obtained \cite{Srinivasan:1998ty}. 
If $i\epsilon$ is introduced with the opposite sign, one obtains
a result which would be regarded as that for the time reversed one
 $\Gamma_{{\rm ab}}=e^{-8\pi \omega M}\Gamma_{{\rm em}}$
for white hole semiclassical absorption rate.
In \cite{Vanzo:2011wq}, 
it is discussed that the $i\epsilon$ prescription
is consistent with the analytic continuation when
spacetime allows the Kruskal extension as in \cite{Hartle:1976tp}.
$i\epsilon$ prescription should correspond to boundary conditions 
on the wave function and choice of vacuum
while in the tunneling method with the WKB approximation identification 
of the positive frequency solutions is usually implicit in the coordination
and in making ansatz for the solution to the Hamilton-Jacobi equation 
\cite{Stotyn:2008qu}.
},
the imaginary part of $S_{out}$ is
\begin{align}
{\rm Im} \; S_{out} = {\rm Im} \int_{a\to b}
\frac{2\omega r^2 dr}
{(r - r_h  - i \e) (r -r_f)}
=
4\pi \omega M,
\label{soutsch}
\end{align}
dropping terms of order $f^2$ or higher.
The semiclassical emission rate is thus obtained as
\be
\Gamma_{\rm em}\propto \exp(-2{\rm Im}S_{out})=\exp(-8\pi \omega M).
\ee
Identifying $\Gamma_{\rm em}$ with the Boltzmann factor, $\exp(-\omega/T)$,
we can read off the Hawking temperature as $T=\frac{\kappa_{s}}{2\pi}$ 
with $\kappa_{s}=\frac{1}{4M}$ the surface gravity
of Schwarzschild black hole. Note that in this computation, only the
infinitesimal region across the trapping 
horizon contributes to the result. 
Note also that the Hawking radiation in this case is independent of
the soft hair features of the black hole.

\subsection{Supertranslated Vaidya Black Hole}

In the above, we find that 
the Hawking radiation is insensitive to the features of soft
hair of a stationary Schwarzschild black hole.
Next, we would like to consider the  process
of black hole evaporation and
compute the Hawking radiation of this dynamical black hole; and
investigate how the soft hair would affect the Hawking radiation.
The black hole evolution  process can be modeled with a supertranslated 
Vaidya spacetime with  a time-dependent
mass $M(v)$ where $v$ is the advanced null coordinate.
The decrease of the black hole mass due to the
ingoing negative energy flux can be interpreted in the tunneling method as 
one having the negative energy particle of a virtual particle pair 
created near the FOTH falls into the black hole. 
Here we effectively incorporate this back reaction effect of ingoing
Hawking
radiation by the diminishing behaviour of the black hole mass $M(v)$
of the background geometry
with $M'<0$ during the evaporation process.

\begin{figure}[t!] 
\centering
\vspace{-1.5cm}
\includegraphics[width=8cm]{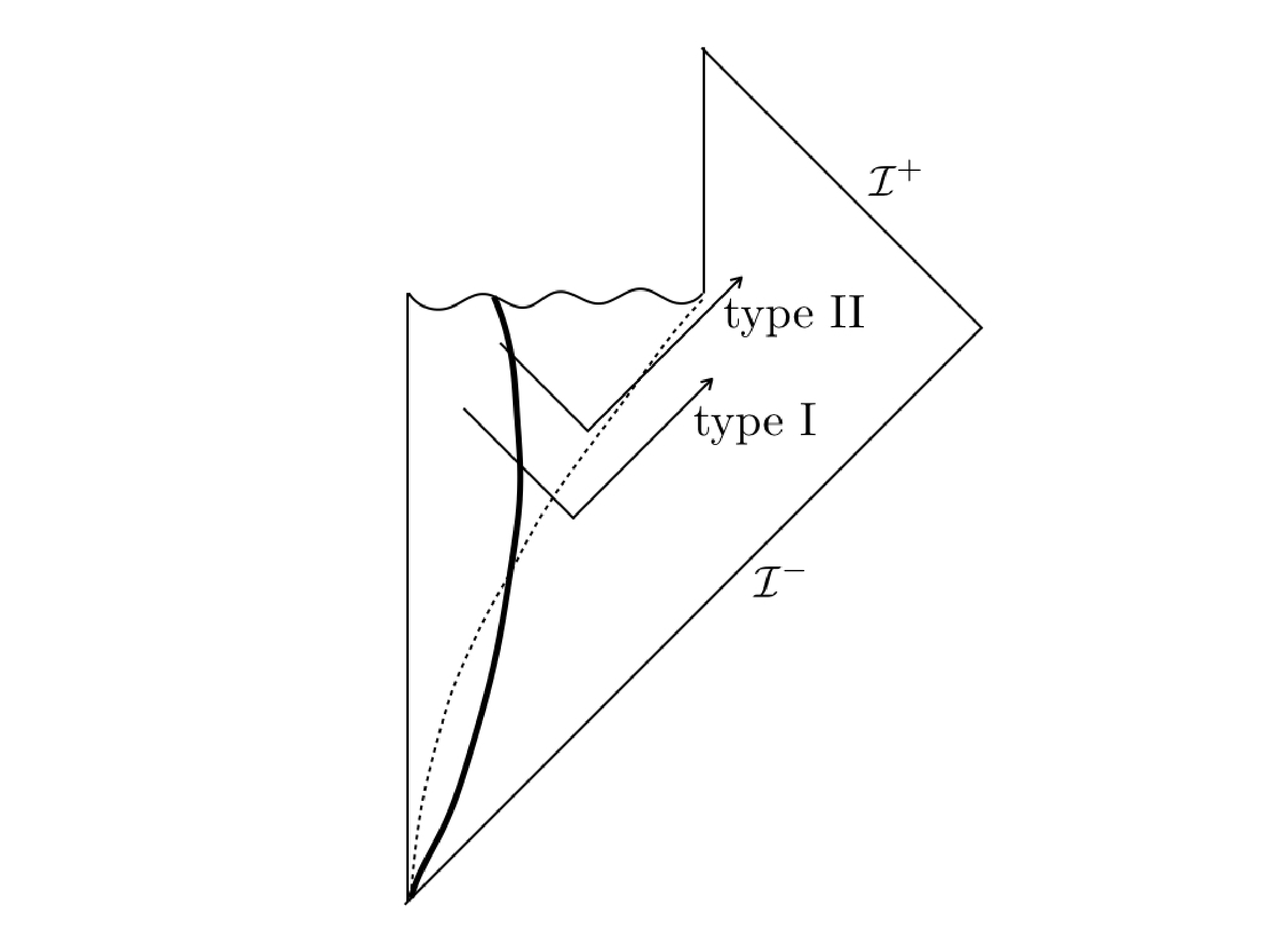}
\vspace{-0.5cm}
\caption{Typical Penrose diagram of the supertranslated Vaidya black hole 
 at a fixed value of $\Theta^{A}$.
The dashed line represents the FOTH $r=r_{h}$, the wavy line represents the singularity $r=0$
and the thick curve represents a collapsing matter.} 
\label{fig:2}
\end{figure}

For the supertranslated Vaidya metric, the Hamilton-Jacobi method
can be applied in a similar way to the previous example of
Schwarzschild black hole.
However there are important differences.
We start with the Hamilton-Jacobi equation \eqref{hjeq} with the metric
given by \eqref{svaidya}.
\begin{align}
2\partial_{r}S\partial_{v}S
&
+
\Big(V-\frac{2fM'}{r}-\frac{MD^{2}f}{r^{2}}\Big)(\partial_{r}S)^{2}
\notag\\
&\hspace{-1.5cm}
+
\frac{1}{r^{2}}D^{A}(2Vf+D^{2}f)\partial_{r}S\partial_{A}S
+
\frac{1}{r^{4}}(r^{2}\gamma^{AB}-2rD^{A}D^{B}f+r\gamma^{AB}D^{2}f)
\partial_{A}S\partial_{B}S
=0.
\label{HJsvaidya}
\end{align}
Here we define the invariant 
energy of particle $\omega$ 
in favor of the Kodama-like vector which generalizes the particle
energy in the static
case. 
\be
\omega \equiv -K^{\mu}p_{\mu}
= -K^{\mu}\partial_{\mu}S=-\partial_{v}S.
\label{energyk}
\ee 
\eqref{HJsvaidya} is rather complicated to solve in general.
The simplest possibility  is to consider  radial null geodesics 
with $\Theta^{A}=\text{const}.$ along the geodesics.
In this case, the Hamilton-Jacobi equation has simple solutions
\begin{align}
\partial_{r}S_{out}
=
2\omega \left(V-\frac{2fM'}{r}-\frac{MD^{2}f}{r^{2}}\right)^{{-1}},
\qquad
\partial_{r}S_{in} = 0.
\label{drS}
\end{align}
And the particle action is reconstructed as
\be \label{sout1}
S_{out}= -\int \omega dv 
+ \int \frac{2\omega r^{2}dr}{Vr^{2}-2fM'r-MD^{2}f},
\qquad
S_{{in}}= -\int \omega dv.
\ee 
As before, $S_{in}$ has no imaginary contribution since the energy is real.
As for $S_{out}$, the line integral contribution \eq{sout1} to
$S_{out}$ is real for
most part of the path except possibly at around the location where the
integrand diverges,
which occurs at the location of the trapping
horizon. However since now the trapping horizon is dynamical, it is no
longer null. 
This leads to two types of tunneling paths,
depending on how the path 
crosses the trapping horizon.
See Fig.\ref{fig:2}.
On the segment of the geodesic which 
crosses the trapping horizon,
they are described by the type-I and type-II paths respectively:
\bea
v= \text{const.}, \qquad && (\text{type-I}),
\label{I}
\\
\frac{dr}{dv} = \frac12
\left(V-\frac{2fM'}{r}-\frac{MD^{2}f}{r^{2}}\right),
\qquad && (\text{type-II}).
\label{II}
\eea
The two types of tunneling paths correspond to
whether the virtual pair is created inside or outside
of the trapping horizon.
The crossing with type-I path corresponds to a pair forming outside and describes
a backward radial null ray 
which
comes out from the
future singularity at $r=0$, crosses the trapping horizon and 
escapes to infinity.
The crossing with type-II path corresponds to a pair forming inside and
describes a backward null ray 
which comes out from the future singularity,
and reaches some interior point in the trapped
region, and then 
crosses the trapping horizon and eventually escapes to infinity. 
The main difference between the two types of tunneling paths is that 
type-I tunneling path crosses the FOTH
along a classically forbidden trajectory backward in time,
while type-II tunneling path crosses the FOTH
along a classically allowed trajectory forward in time.
Note that the crossing with type-II path is absent for a static black hole.

Let us now compute ${\rm Im} S_{out}$.  It is clear that $S_{out}$
vanishes and no tunneling occurs for type-II paths.
For type-I paths, we have \eq{S-out} 
with $r_{h}$ given by \eqref{rh}
and it is clear that the integrand has a pole on
the FOTH $r=r_{h}$.
The imaginary contribution due to the pole is evaluated 
by an $i\epsilon$ prescription as before and yields
\begin{align}
  {\rm Im}S_{out}
=
\frac{2\pi \omega r_{h}^{2}}
{r_{h}-r_{f}}
=
\frac{\pi \omega}{\kappa},
\label{imssvaidya}
\end{align} 
where $\omega$ 
is evaluated on the FOTH for a fixed $v$ and $\Theta^{A}$.
In the last equality of \eqref{imssvaidya}, we have used
\eqref{sgr2}  to write the result in terms of 
the horizon surface gravity $\kappa$,
\be
\label{kappa}
\kappa =
\frac{r_{h}-r_{f}}{r_{h}^{2}}
=
\frac{1}{4M}\Big(1-f\frac{M'}{M}\Big).
\ee
The semiclassical emission rate $\Gamma_{{\rm em}}$ 
is thus formally written in the same form as that in the static case.
\be
\Gamma_{{\rm em}} \propto e^{-2 {\rm Im}S_{out}}
= \exp\left(- \frac{2\pi\omega}{\kappa}\right).
\label{emsv}
\ee
%
We find that the supertranslation $f(\Theta)$ dependence comes into 
the spectrum of Hawking radiation through the surface gravity $\kappa$ 
of the FOTH.  

Note that our result \eq{emsv} of the emission rate  is
consistent with the factorization property of the black
hole S-matrix
\cite{
  Mirbabayi:2016axw, Gabai:2016kuf,Gomez:2016hxz,Bousso:2017dny}.
The non-trivial dependence on the soft hair in
\eq{emsv} is due to the fact that we are performing an
observation of the undressed hard Hawking radiation quanta. If one is to dress
the hard modes with soft gravitons, then the soft hair dependence
would disappear. This is similar to the discussion of gravitational
memory where one may also remove the memory with dressing
\cite{Bousso:2017dny}. It should be clear that the factorization of soft modes and
hard modes in the S-matrix does not imply that the soft hair has no
physical implication at all.  The effects of soft hair on
gravitational memory and Hawking radiation spectrum is physical and is
observable.

So far we have left $M'$ free and 
determined the Hawking radiation in
terms of it. In a consistent model of evaporation, the mass loss
\cite{Hawking:1974sw,Page:1976df} is
caused by the Hawking radiation.
The leading contribution to the mass loss of the
supertranslated Vaidya black hole 
can be estimated from the Hawking radiation spectrum as:
\begin{align}
-M'
&=
\frac{1}{2\pi^{2}}\int 
\left(\int^{\infty}_{0} \frac{\omega^{3}}{e^{\beta \omega }-1} d\omega \right) 
\; r_{h}^{2} \sqrt{\gamma}d^2{\Theta}
= 
\frac{\pi^{2}}{30}\int \frac{r_{h}^{2}}{\beta^{4}} \sqrt{\gamma}d^2{\Theta}
 %
\notag\\
&=
\Big(1-\frac{a_{00}}{\sqrt{\pi}}\frac{M'}{M}\Big){P_0},
\label{mloss}
\end{align}
where $\beta:=2\pi/\kappa$ and
${P_0}:=(7680\pi M^{2})^{-1}$ is the
standard power loss due to  Hawking radiation in the leading order
approximation of a constant $M$.
Note that 
non-trivial angle-dependent supertranslations 
corresponding to $l\geq 1$ do not contribute to the mass loss
in the case of supertranslated Vadiya black hole with $M =M(v)$.
%
Solving \eqref{mloss} for $M'$ we get
\be
M'= -{ P_0}\Big(1-\frac{a_{00}{P_0}}{\sqrt{\pi}M} \Big)^{-1}.
\ee
With this value of $M'$, we obtain a consistent model of dynamical
black hole whose evaporation is driven by the power loss due to
Hawking radiation.

Let us discuss the entropy of the supertranslated Vaidya black hole.
A proposal for dynamical black hole entropy is presented by
Iyer and Wald \cite{Iyer:1994ys}.
For Einstein gravity with matter minimally coupled to gravity, 
the entropy of dynamical black hole is given by the same formula
$S_{\rm dyn}= {A_{h}}/{4}$ as for
stationary black holes. We find
\begin{align}
S_{\rm dyn}
&=
\frac{1}{4}
\int 
r_{h}^{2}
\sqrt{\gamma}d^{2}\Theta
\notag\\
&
=
4\pi M^{2} + 4\sqrt{\pi}  a_{00} M M'.
\label{sdynamical}
\end{align}
The entropy \eq{sdynamical} can be expressed as 
$S_{\rm dyn} = 4 \pi
M^2(v+a_{00}/\sqrt{4\pi})$ 
and it is consistent with the fact that only the zero
mode part of the shift of time can appear in the mass function.

When the black hole emits a massless Hawking radiation,
the black hole mass changes
by amount of $\omega$, $M \to M-\omega$ 
due to the energy conservation,
and the entropy changes accordingly
$S_{\rm dyn}
\to
S_{\rm dyn}+\Delta S$
with
\be
\Delta S
=
- \omega(8\pi M + 4\sqrt{\pi} a_{00} M') + {\cal O}(\omega^{2}).
\ee
This can be related to the
semiclassical emission rate \eqref{emsv} as
\be \label{SG}
\Delta S = 
\int \frac{d\Omega}{4 \pi} \ln \G_{\rm em}.
\ee
This generalizes the standard relation for spherical black holes
\cite{Parikh:1999mf,Vanzo:2011wq}. 
Furthermore,  one may define a differential
entropy change:
\be \label{SG1}
\frac{d \Delta S}{d \Omega} = 
 \ln \G_{\rm em}
\ee
and
interpret the relation \eq{SG} as saying that the
Hawking radiation carries away from the black hole 
different amount of entropy \eq{SG1}
at different angles. As Hawking
radiation originates locally at the surface of the trapping horizon,
it is consistent that a local change of entropy of the black hole occurs.

\section{Conclusion and Discussion}

Supertranslation of black hole adds soft hair to it. These soft hairs
are physical and can be observed through the classical memory
effect. In this paper, we show that  soft hairs also have non-trivial
effects on the quantum physics of black hole. In particular we computed
the Hawking radiation for a dynamical black hole modeled by the
Vaidya spacetime with soft hair. We find that tunneling occurs at the 
trapping horizon and the semiclassical emission rate of Hawking
radiation is characterized by 
the horizon surface gravity $\kappa$ defined by 
the Kodama-like vector
of the supertranslated Vaidya black hole.
The Hawking radiation spectrum has a dependence on the soft hair
distribution over the black hole.
Our results make it clear that soft hair of black hole is physical
and it has clear observable effects on the Hawking radiation spectrum.
Of course this depends on what is being observed. 
If one wishes, one may
also choose to make detection of the dressed Hawking
quanta, then the observed spectrum would become the canonical 
one without any sight of the soft hair. 

Recently Strominger
\cite{Strominger:2017aeh} has emphasized the presence of soft modes in
the final state of the black hole evaporation and 
proposed that the final state of the black hole evaporation process
to be 
a pure state of the form
\be \label{final-state}
\ket{\Psi} = \sum_\a c_\a \ket{H_\a} \ket{S_\a},
\ee
where  $\ket{H_\a}$ describes the state of radiation in the
thermal ensemble of Hawking radiation and $\ket{S_\a}$ describes the
cloud of soft modes that is required by charge conservation.
In other words, the purity of the quantum state
is restored by the soft graviton modes.
However if only
the hard modes of the Hawking radiation are to be observed, the resulting reduced
density matrix
$\rho_r = \tr_{\rm soft} |\Psi \rangle \langle \Psi | = \rho_{\rm thermal}$
  gives the thermal density matrix of Hawking radiation. Our result is
  consistent with this.
  The loss of coherence when ``environmental variables'' are traced
  out is not special and can be expected generically.
  Indeed, similar
  decoherence effect
  of hard particles as a result of the tracing out
  of unobservable soft modes has been demonstrated for certain kind of
  events in QED  \cite{Carney:2017jut,Carney:2017oxp}. 
It is important
  to understand deeper how \eq{final-state} is related to
  the BMS symmetry and the infrared structure of 
  gravity.


\vskip7mm
\section*{Acknowledgements}

We would like to thank Carlos Cardona, 
Dimitrios Giataganas, Pei-Ming Ho, Yoshinori Honma, Yu-tin Huang, 
Janet Hung, Satoshi Iso, Shoichi Kawamoto, Hiroyuki Kitamoto
Chris Lau, Misao Sasaki, Jiro Soda and Wei Song for 
useful discussions and comments. We thank in particular Misao Sasaki
for reading the manuscript and for valuable comments and suggestions.
We benefited much from the discussions on BMS symmetries and soft hair
at KEK, Japan, during the conference 
``East Asia Joint Workshop on Fields and Strings 2017''
 held in November 13-17, 2017, 
 and also at Yukawa Institute for Theoretical Physics, Japan,
 during the conference ``CosPA2017'' held in December 11-15, 2017. 
This work is
supported in part by the National Center of Theoretical Science
(NCTS) and the grant
104-2112-M-007-001-MY3 of the Ministry of Science and
Technology of Taiwan.


\appendix




\begin{thebibliography}{100}


\bibitem{Bekenstein:1973ur}
  J.~D.~Bekenstein,
  ``Black holes and entropy,''
  Phys.\ Rev.\ D {\bf 7} (1973) 2333.


\bibitem{Hawking:1974rv} 
  S.~W.~Hawking,
  ``Black hole explosions,''
  Nature {\bf 248}, 30 (1974).
  
\bibitem{Hawking:1974sw}
  S.~W.~Hawking,
  ``Particle Creation by Black Holes,''
  Commun.\ Math.\ Phys.\  {\bf 43} (1975) 199
   Erratum: [Commun.\ Math.\ Phys.\  {\bf 46} (1976) 206].

  
\bibitem{Wald:1993nt}
  R.~M.~Wald,
  ``Black hole entropy is the Noether charge,''
  Phys.\ Rev.\ D {\bf 48}, no. 8, R3427 (1993)
  [gr-qc/9307038].
  

\bibitem{Chrusciel:2012jk}
  P.~T.~Chrusciel, J.~Lopes Costa and M.~Heusler,
  ``Stationary Black Holes: Uniqueness and Beyond,''
  Living Rev.\ Rel.\  {\bf 15} (2012) 7
  [arXiv:1205.6112 [gr-qc]].

\bibitem{Mathur:2009hf}
  S.~D.~Mathur,
  ``The Information paradox: A Pedagogical introduction,''
  Class.\ Quant.\ Grav.\  {\bf 26} (2009) 224001
  [arXiv:0909.1038 [hep-th]].

\bibitem{Marolf:2017jkr}
  D.~Marolf,
  ``The Black Hole information problem: past, present, and future,''
  Rept.\ Prog.\ Phys.\  {\bf 80} (2017) no.9,  092001
  [arXiv:1703.02143 [gr-qc]].
  
  
\bibitem{Strominger:2013jfa} 
  A.~Strominger,
  ``On BMS Invariance of Gravitational Scattering,''
  JHEP {\bf 1407}, 152 (2014)
  [arXiv:1312.2229 [hep-th]].


\bibitem{He:2014laa} 
  T.~He, V.~Lysov, P.~Mitra and A.~Strominger,
  ``BMS supertranslations and Weinberg's soft graviton theorem,''
  JHEP {\bf 1505}, 151 (2015)
  [arXiv:1401.7026 [hep-th]].

\bibitem{Cachazo:2014fwa}
  F.~Cachazo and A.~Strominger,
  ``Evidence for a New Soft Graviton Theorem,''
  arXiv:1404.4091 [hep-th].
  
 \bibitem{Kapec:2014opa}
  D.~Kapec, V.~Lysov, S.~Pasterski and A.~Strominger,
  ``Semiclassical Virasoro symmetry of the quantum gravity $ \mathcal{S}$-matrix,''
  JHEP {\bf 1408} (2014) 058
  [arXiv:1406.3312 [hep-th]].


\bibitem{Strominger:2014pwa} 
  A.~Strominger and A.~Zhiboedov,
  ``Gravitational Memory, BMS Supertranslations and Soft Theorems,''
  JHEP {\bf 1601}, 086 (2016)
  [arXiv:1411.5745 [hep-th]].


  
\bibitem{Strominger:2017zoo} 
  A.~Strominger,
  ``Lectures on the Infrared Structure of Gravity and Gauge Theory,''
  arXiv:1703.05448 [hep-th].

  
\bibitem{Hawking:2016msc} 
  S.~W.~Hawking, M.~J.~Perry and A.~Strominger,
  ``Soft Hair on Black Holes,''
  Phys.\ Rev.\ Lett.\  {\bf 116}, no. 23, 231301 (2016)
  [arXiv:1601.00921 [hep-th]].


\bibitem{Hawking:2016sgy} 
  S.~W.~Hawking, M.~J.~Perry and A.~Strominger,
  ``Superrotation Charge and Supertranslation Hair on Black Holes,''
  JHEP {\bf 1705}, 161 (2017)
  [arXiv:1611.09175 [hep-th]].
  
\bibitem{Strominger:2017aeh}
  A.~Strominger,
  ``Black Hole Information Revisited,''
  arXiv:1706.07143 [hep-th].

  
\bibitem{Bondi:1962px} 
  H.~Bondi, M.~G.~J.~van der Burg and A.~W.~K.~Metzner,
  ``Gravitational waves in general relativity. 7. Waves from axisymmetric isolated systems,''
  Proc.\ Roy.\ Soc.\ Lond.\ A {\bf 269}, 21 (1962).

\bibitem{Sachs:1962wk} 
  R.~K.~Sachs,
  ``Gravitational waves in general relativity. 8. Waves in asymptotically flat space-times,''
  Proc.\ Roy.\ Soc.\ Lond.\ A {\bf 270}, 103 (1962).

\bibitem{Arnowitt:1962hi} 
  R.~L.~Arnowitt, S.~Deser and C.~W.~Misner,
  ``The Dynamics of general relativity,''
  Gen.\ Rel.\ Grav.\  {\bf 40}, 1997 (2008)
  [gr-qc/0405109].

\bibitem{Wald:1999wa} 
  R.~M.~Wald and A.~Zoupas,
  ``A General definition of 'conserved quantities' in general relativity and other theories of gravity,''
  Phys.\ Rev.\ D {\bf 61}, 084027 (2000)
  [gr-qc/9911095].

\bibitem{Barnich:2011mi} 
  G.~Barnich and C.~Troessaert,
  ``BMS charge algebra,''
  JHEP {\bf 1112}, 105 (2011)
  [arXiv:1106.0213 [hep-th]].


\bibitem{Zeldovich} 
Ya. B. Zeldovich and A. G. Polnarev,
``Radiation of gravitational waves by a cluster of superdense stars,'' 
Sov.\ Astron.\ {\bf 18}, 17 (1974).

\bibitem{Braginsky:1986ia} 
  V.~B.~Braginsky and L.~P.~Grishchuk,
  ``Kinematic Resonance and Memory Effect in Free Mass Gravitational Antennas,''
  Sov.\ Phys.\ JETP {\bf 62}, 427 (1985)
  [Zh.\ Eksp.\ Teor.\ Fiz.\  {\bf 89}, 744 (1985)].
   
\bibitem{Braginskii:1987} 
V.~B.~Braginsky and K.~S.~Thorne,
``Gravitational-wave bursts with memory and experimental prospects,''
 Nature, {\bf 327}, 123 (1987).
 
\bibitem{Blanchet:1992br} 
  L.~Blanchet and T.~Damour,
  ``Hereditary effects in gravitational radiation,''
  Phys.\ Rev.\ D {\bf 46}, 4304 (1992).
  
  
\bibitem{Mirbabayi:2016axw} 
  M.~Mirbabayi and M.~Porrati,
  ``Dressed Hard States and Black Hole Soft Hair,''
  Phys.\ Rev.\ Lett.\  {\bf 117}, no. 21, 211301 (2016)
  [arXiv:1607.03120 [hep-th]].
  
\bibitem{Gabai:2016kuf}
  B.~Gabai and A.~Sever,
  ``Large gauge symmetries and asymptotic states in QED,''
  JHEP {\bf 1612} (2016) 095
  [arXiv:1607.08599 [hep-th]].

\bibitem{Gomez:2016hxz}
  C.~Gomez and M.~Panchenko,
  ``Asymptotic dynamics, large gauge transformations and infrared symmetries,''
  arXiv:1608.05630 [hep-th].
  
\bibitem{Bousso:2017dny} 
  R.~Bousso and M.~Porrati,
  ``Soft Hair as a Soft Wig,''
  Class.\ Quant.\ Grav.\  {\bf 34}, no. 20, 204001 (2017)
  [arXiv:1706.00436 [hep-th]].
  

\bibitem{Donnelly:2017jcd}
  W.~Donnelly and S.~B.~Giddings,
  ``How is quantum information localized in gravity?,''
  Phys.\ Rev.\ D {\bf 96} (2017) no.8,  086013
  [arXiv:1706.03104 [hep-th]].


\bibitem{Srinivasan:1998ty} 
  K.~Srinivasan and T.~Padmanabhan,
  ``Particle production and complex path analysis,''
  Phys.\ Rev.\ D {\bf 60}, 024007 (1999)
  [gr-qc/9812028].

  
\bibitem{Parikh:1999mf} 
  M.~K.~Parikh and F.~Wilczek,
  ``Hawking radiation as tunneling,''
  Phys.\ Rev.\ Lett.\  {\bf 85}, 5042 (2000)
  [hep-th/9907001].
  
  
 
 
     
\bibitem{Koga:2001vq} 
  J.~i.~Koga,
  ``Asymptotic symmetries on Killing horizons,''
  Phys.\ Rev.\ D {\bf 64}, 124012 (2001)
  [gr-qc/0107096].
 
\bibitem{Hotta:2000gx} 
  M.~Hotta, K.~Sasaki and T.~Sasaki,
  ``Diffeomorphism on horizon as an asymptotic isometry of Schwarzschild black hole,''
  Class.\ Quant.\ Grav.\  {\bf 18}, 1823 (2001)
  [gr-qc/0011043].
  
 
\bibitem{Donnay:2015abr} 
  L.~Donnay, G.~Giribet, H.~A.~Gonzalez and M.~Pino,
  ``Supertranslations and Superrotations at the Black Hole Horizon,''
  Phys.\ Rev.\ Lett.\  {\bf 116}, no. 9, 091101 (2016)
  [arXiv:1511.08687 [hep-th]].
     
\bibitem{Averin:2016ybl} 
  A.~Averin, G.~Dvali, C.~Gomez and D.~Lust,
  ``Gravitational Black Hole Hair from Event Horizon Supertranslations,''
  JHEP {\bf 1606}, 088 (2016)
  [arXiv:1601.03725 [hep-th]].
  
\bibitem{Eling:2016xlx} 
  C.~Eling and Y.~Oz,
  ``On the Membrane Paradigm and Spontaneous Breaking of Horizon BMS Symmetries,''
  JHEP {\bf 1607}, 065 (2016)
  [arXiv:1605.00183 [hep-th]].
  
  
\bibitem{Donnay:2016ejv} 
  L.~Donnay, G.~Giribet, H.~A.~Gonzalez and M.~Pino,
  ``Extended Symmetries at the Black Hole Horizon,''
  JHEP {\bf 1609}, 100 (2016)
  [arXiv:1607.05703 [hep-th]].
  
\bibitem{Cai:2016idg} 
  R.~G.~Cai, S.~M.~Ruan and Y.~L.~Zhang,
  ``Horizon supertranslation and degenerate black hole solutions,''
  JHEP {\bf 1609}, 163 (2016)
  [arXiv:1609.01056 [gr-qc]].
  
\bibitem{Afshar:2016kjj} 
  H.~Afshar, D.~Grumiller, W.~Merbis, A.~Perez, D.~Tempo and R.~Troncoso,
  ``Soft hairy horizons in three spacetime dimensions,''
  Phys.\ Rev.\ D {\bf 95}, no. 10, 106005 (2017)
  [arXiv:1611.09783 [hep-th]].
  
    
  
 
\bibitem{Hotta:2016qtv} 
  M.~Hotta, J.~Trevison and K.~Yamaguchi,
  ``Gravitational Memory Charges of Supertranslation and Superrotation on Rindler Horizons,''
  Phys.\ Rev.\ D {\bf 94}, no. 8, 083001 (2016)
  [arXiv:1606.02443 [gr-qc]].
  
\bibitem{Kolekar:2017yoi} 
  S.~Kolekar and J.~Louko,
  ``Gravitational memory for uniformly accelerated observers,''
  Phys.\ Rev.\ D {\bf 96}, no. 2, 024054 (2017)
  [arXiv:1703.10619 [hep-th]].
    
\bibitem{Kolekar:2017tge} 
  S.~Kolekar and J.~Louko,
  ``Quantum memory for Rindler supertranslations,''
  arXiv:1709.07355 [hep-th].
  
  
  
\bibitem{Barnich:2009se} 
  G.~Barnich and C.~Troessaert,
  ``Symmetries of asymptotically flat 4 dimensional spacetimes at null infinity revisited,''
  Phys.\ Rev.\ Lett.\  {\bf 105}, 111103 (2010)
  [arXiv:0909.2617 [gr-qc]].
 

\bibitem{Sachs:1962zza} 
  R.~Sachs,
  ``Asymptotic symmetries in gravitational theory,''
  Phys.\ Rev.\  {\bf 128}, 2851 (1962).


  
\bibitem{Hayward:1993wb} 
  S.~A.~Hayward,
  ``General laws of black hole dynamics,''
  Phys.\ Rev.\ D {\bf 49}, 6467 (1994).


\bibitem{Ashtekar:1998sp} 
  A.~Ashtekar, C.~Beetle and S.~Fairhurst,
  ``Isolated horizons: A Generalization of black hole mechanics,''
  Class.\ Quant.\ Grav.\  {\bf 16}, L1 (1999)
  [gr-qc/9812065].

\bibitem{Ashtekar:2002ag} 
  A.~Ashtekar and B.~Krishnan,
  ``Dynamical horizons: Energy, angular momentum, fluxes and balance laws,''
  Phys.\ Rev.\ Lett.\  {\bf 89}, 261101 (2002)
  [gr-qc/0207080].
  
\bibitem{Booth:2003ji} 
  I.~Booth and S.~Fairhurst,
  ``The First law for slowly evolving horizons,''
  Phys.\ Rev.\ Lett.\  {\bf 92}, 011102 (2004)
  [gr-qc/0307087].

\bibitem{Ashtekar:2004cn} 
  A.~Ashtekar and B.~Krishnan,
  ``Isolated and dynamical horizons and their applications,''
  Living Rev.\ Rel.\  {\bf 7}, 10 (2004)
  [gr-qc/0407042].


\bibitem{Booth:2005qc} 
  I.~Booth,
  ``Black hole boundaries,''
  Can.\ J.\ Phys.\  {\bf 83}, 1073 (2005)
  [gr-qc/0508107].

\bibitem{Hayward:2008ti} 
  S.~A.~Hayward,
  ``Dynamics of black holes,''
  arXiv:0810.0923 [gr-qc].

\bibitem{Gourgoulhon:2008pu} 
  E.~Gourgoulhon and J.~L.~Jaramillo,
  ``New theoretical approaches to black holes,''
  New Astron.\ Rev.\  {\bf 51}, 791 (2008)
  [arXiv:0803.2944 [astro-ph]].

\bibitem{Faraoni:2013aba} 
  V.~Faraoni,
  ``Evolving black hole horizons in General Relativity and alternative gravity,''
  Galaxies {\bf 1}, no. 3, 114 (2013)
  [arXiv:1309.4915 [gr-qc]].
  
  
\bibitem{Wald:1984rg} 
  R.~M.~Wald,
  ``General Relativity,''
  University of Chicago Press (1984) 491p.
  
  
\bibitem{Nielsen:2007ac} 
  A.~B.~Nielsen and J.~H.~Yoon,
  ``Dynamical surface gravity,''
  Class.\ Quant.\ Grav.\  {\bf 25}, 085010 (2008)
  [arXiv:0711.1445 [gr-qc]].


\bibitem{Vanzo:2011wq} 
  L.~Vanzo, G.~Acquaviva and R.~Di Criscienzo,
  ``Tunnelling Methods and Hawking's radiation: achievements and prospects,''
  Class.\ Quant.\ Grav.\  {\bf 28}, 183001 (2011)
  [arXiv:1106.4153 [gr-qc]].


\bibitem{Kodama:1979vn} 
  H.~Kodama,
  ``Conserved Energy Flux for the Spherically Symmetric System and the Back Reaction Problem in the Black Hole Evaporation,''
  Prog.\ Theor.\ Phys.\  {\bf 63}, 1217 (1980).


\bibitem{Hayward:1997jp} 
  S.~A.~Hayward,
  ``Unified first law of black hole dynamics and relativistic thermodynamics,''
  Class.\ Quant.\ Grav.\  {\bf 15}, 3147 (1998)
  [gr-qc/9710089].

\bibitem{Kerner:2006vu} 
  R.~Kerner and R.~B.~Mann,
  ``Tunnelling, temperature and Taub-NUT black holes,''
  Phys.\ Rev.\ D {\bf 73}, 104010 (2006)
  [gr-qc/0603019].

\bibitem{Visser:2001kq} 
  M.~Visser,
  ``Essential and inessential features of Hawking radiation,''
  Int.\ J.\ Mod.\ Phys.\ D {\bf 12}, 649 (2003)
  [hep-th/0106111].
  
\bibitem{DiCriscienzo:2007pcr} 
  R.~Di Criscienzo, M.~Nadalini, L.~Vanzo, S.~Zerbini and G.~Zoccatelli,
  ``On the Hawking radiation as tunneling for a class of dynamical black holes,''
  Phys.\ Lett.\ B {\bf 657}, 107 (2007)
  [arXiv:0707.4425 [hep-th]].
  
\bibitem{Hayward:2008jq} 
  S.~A.~Hayward, R.~Di Criscienzo, L.~Vanzo, M.~Nadalini and S.~Zerbini,
  ``Local Hawking temperature for dynamical black holes,''
  Class.\ Quant.\ Grav.\  {\bf 26}, 062001 (2009)
  [arXiv:0806.0014 [gr-qc]].

\bibitem{Hartle:1976tp} 
  J.~B.~Hartle and S.~W.~Hawking,
  ``Path Integral Derivation of Black Hole Radiance,''
  Phys.\ Rev.\ D {\bf 13}, 2188 (1976).

\bibitem{Chatterjee:2007hc} 
  B.~Chatterjee, A.~Ghosh and P.~Mitra,
  ``Tunnelling from black holes in the Hamilton Jacobi approach,''
  Phys.\ Lett.\ B {\bf 661}, 307 (2008)
  [arXiv:0704.1746 [hep-th]].


 
\bibitem{Stotyn:2008qu} 
  S.~Stotyn, K.~Schleich and D.~Witt,
  ``Observer Dependent Horizon Temperatures: A Coordinate-Free Formulation of Hawking Radiation as Tunneling,''
  Class.\ Quant.\ Grav.\  {\bf 26}, 065010 (2009)
  [arXiv:0809.5093 [gr-qc]].
  
 
\bibitem{Page:1976df} 
  D.~N.~Page,
  ``Particle Emission Rates from a Black Hole: Massless Particles from an Uncharged, Nonrotating Hole,''
  Phys.\ Rev.\ D {\bf 13}, 198 (1976).
  
  
\bibitem{Iyer:1994ys} 
  V.~Iyer and R.~M.~Wald,
  ``Some properties of Noether charge and a proposal for dynamical black hole entropy,''
  Phys.\ Rev.\ D {\bf 50}, 846 (1994)
  [gr-qc/9403028].
 

 
 
\bibitem{Carney:2017jut}
  D.~Carney, L.~Chaurette, D.~Neuenfeld and G.~W.~Semenoff,
  ``Infrared quantum information,''
  Phys.\ Rev.\ Lett.\  {\bf 119} (2017) no.18,  180502
  [arXiv:1706.03782 [hep-th]].

\bibitem{Carney:2017oxp}
  D.~Carney, L.~Chaurette, D.~Neuenfeld and G.~W.~Semenoff,
  ``Dressed infrared quantum information,''
  arXiv:1710.02531 [hep-th].



  
\end{thebibliography}
 \end{document}